\documentclass[10pt, twocolumn]{manuscript} 

\usepackage{color}
\definecolor{darkblue}{rgb}{0.0, 0.0, 0.4}
\definecolor{darkred}{rgb}{0.8, 0.0, 0.0}
\definecolor{darkgreen}{rgb}{0.0, 0.7, 0.0}
\definecolor{purple}{rgb}{0.4, 0.1, 0.8}
\definecolor{medblue}{rgb}{0.05, 0.45, 0.7}

\usepackage{hyperref}
\hypersetup{
    colorlinks=true,
    linkbordercolor={white},
    linkcolor=darkred,
    citecolor=medblue,
    urlcolor=blue
}

\usepackage[nameinlink,noabbrev]{cleveref}

\usepackage{caption,subcaption}
\usepackage{mathrsfs}
\DeclareMathAlphabet{\mathscr}{OT1}{pzc}{m}{it}
\usepackage{lipsum}
\usepackage{adjustbox}  

\usepackage{url}
\usepackage{doi}
\usepackage{float}

\usepackage{booktabs}
\usepackage{array}  

\newcommand{\vect}[1]{\mathbf{#1}}

\newcommand{\timewindow}{\omega}
\newcommand{\ubar}[1]{\text{\b{$#1$}}}

\DeclareMathOperator{\atantwo}{atan2}

\newcommand{\beginsupplement}{%
        \setcounter{table}{0}
        \renewcommand{\thetable}{S\arabic{table}}%
        \setcounter{figure}{0}
        \renewcommand{\thefigure}{S\arabic{figure}}%
     }

\usepackage{pgffor} 

\usepackage{siunitx}

\newcommand{\createsubpanels}[2]{%
    \renewcommand{\thesubfigure}{\alph{subfigure}} 
    \setcounter{subfigure}{0} 
    \foreach \n in {1,...,#2} {%
        \begin{subfigure}[b]{0.45\textwidth} 
            \phantomsubcaption
            \label{fig:#1_\thesubfigure} 
        \end{subfigure}
        \hfill
    }
}

\newcommand{\mathcenter}{\@fleqnfalse}

\usepackage[font=small,labelfont=bf]{caption}

\title{Discovering and exploiting active sensing motifs for estimation}

\author[1]{Benjamin Cellini}
\author[1]{Burak Boyacıoğlu}
\author[1]{Austin P. Lopez}
\author[1]{Floris van Breugel}

\affil[1]{Dept. of Mechanical Engineering, University of Nevada, Reno}

\corrauthor{Benjamin Cellini, bcellini00@gmail.com \\ Floris van Breugel, fvanbreugel@unr.edu}


\begin{abstract}
From organisms to machines, autonomous systems rely on measured sensory cues to estimate unknown information about themselves or their environment. For nonlinear systems, strategic sensor motion can be leveraged to extract otherwise inaccessible information. This principle, known as active sensing, is widespread in biology yet difficult to study, and remains underutilized in engineered systems due to the challenge of systematically designing active sensing motifs. Here, we introduce the method ``BOUNDS: Bounding Observability for Uncertain Nonlinear Dynamic Systems", and Python package \textit{pybounds}, which can discover movement motifs that increase the information encoded in sensory cues. To exploit sporadic estimates from bouts of active sensing, we further introduce the Augmented Information Kalman Filter (AI-KF). The AI-KF uses insight from BOUNDS to dynamically fuse neural network and model-based estimation. We demonstrate BOUNDS and the AI-KF on a flying agent model and experimental GPS-denied data from a quadcopter, revealing how specific active movements improve estimates of ground speed, altitude, and wind direction. Altogether, our work will prove useful for designing sensor-minimal autonomous systems and investigating active sensing in living organisms.

\end{abstract}

\begin{document}

\flushbottom
\maketitle
\thispagestyle{empty}

\section*{Introduction}

A fundamental challenge for autonomous agents---organisms \cite{Friston2010} or machines \citep{Thrun2005}---is to combine sensory cues into estimates of task-relevant variables \citep{yang2016theoretical}. This challenge intensifies when variables of interest are nonlinearly related to sensor measurements \citep{sontag2022observabilityresultrelatedactive}. For example, optic flow encodes the ratio of velocity-to-distance, making it impossible to estimate either quantity independently from any single moment \citep{Martinelli2012, VanBreugel2014}. For moving agents with nonlinear dynamics or measurements, strategically applied movement can decouple correlated variables and enhance estimation, a process we refer to as \textit{active sensing} \citep{zweifel2020defining}. By accelerating their vision systems, animals like mice \citep{Parker2022} and mantids \citep{Poteser1995} can gauge distance, while visual-inertial sensor fusion in robots requires similar motion to estimate position \citep{bouazza2025observer, qin2018vins}.

For complex estimation tasks, a major barrier to designing active sensing strategies for machines, and understanding their role in organisms, is determining which movements decouple key variables. Existing approaches leverage the control-theoretic concept \textit{observability}. Closely related to Fisher information \citep{Boyacoglu2024}, observability describes how well system states can be inferred from measurements and inputs \citep{hermann2003nonlinear}. Analytic observability tools have yielded \textit{qualitative} insight into how movement motifs in fish \citep{Stamper2012,Kunapareddy2018}, flies \citep{VanBreugel2021a}, and machines \citep{hinson2013path} can aid in estimation, but their inability to distinguish weak vs. strong observability levels limits their utility, especially in data-driven applications. Current \textit{quantitative} methods, on the other hand, rely on eigenvalues of the observability Gramian \citep{marques1986relative, Krener2009}, making it difficult to evaluate observability of individual state variables for partially observable nonlinear systems \citep{Cellini2023}. Neither accounts for unknown system inputs, preventing their application to experimental trajectory data. Consequently, no existing approach can quantify observability for individual state variables in partially observable nonlinear systems from experimental data, precisely what is needed to systematically discover active sensing motifs for real world applications.

A second challenge is integrating active sensing into state estimation. Classical estimators like the Kalman filter (KF) framework \citep{urrea2021kalman} do not explicitly account for time-varying observability, and can exhibit instability in weakly observable dimensions \citep{Butcher2017, Bocquet2017}. The critical flaw of iterative schemes like the KF is that measurement information is only correctly incorporated when state estimates are already accurate. When estimates are poor, information is effectively lost. Data-driven approaches that can access a history of information all at once offer a potential resolution to this challenge. Increasing interest in artificial neural networks (ANNs) has spawned numerous ANN-KF hybrids \citep{Bai2023} that replace KF components \citep{Li2021,Choi2023,Revach2022}, entire filters \citep{Ghosh2024,krishnan2015deep}, or precondition inputs \citep{Liu2022,Weiss2019}. None, however, provide a theoretically grounded approach for determining \textit{when} and \textit{how much} to weigh model-based predictions with data-driven estimates that draw on a time history of measurements.

Here we introduce BOUNDS (Bounding Observability for Uncertain Nonlinear Dynamic Systems)\footnote{An early version appears in \citep{cellini2024discovering}, which was never submitted for publication.}, an empirical computational pipeline that quantifies observability of individual state variables along dynamic trajectories while accounting for sensor noise through Fisher information and the Cramér-Rao bound \citep{Cramer1946}. BOUNDS reveals which states are estimable, which movement motifs are helpful, and which sensors are required---both for measured trajectories and robotic system design. We further present the Augmented Information Kalman Filter (AI-KF), which provides a framework for dealing with sporadic estimates from active sensing systems. The AI-KF combines model-based filtering with ANN-based estimation by using BOUNDS-derived variance estimates to merge sporadic active sensing information in a statistically meaningful manner. We demonstrate this framework on a flying agent estimating altitude, ground speed, and wind direction with limited sensors, showing the AI-KF outperforms traditional approaches under poor initialization, unmodeled disturbances, and intermittent observability.

\begin{figure*}[!th]
    \centerline{\includegraphics{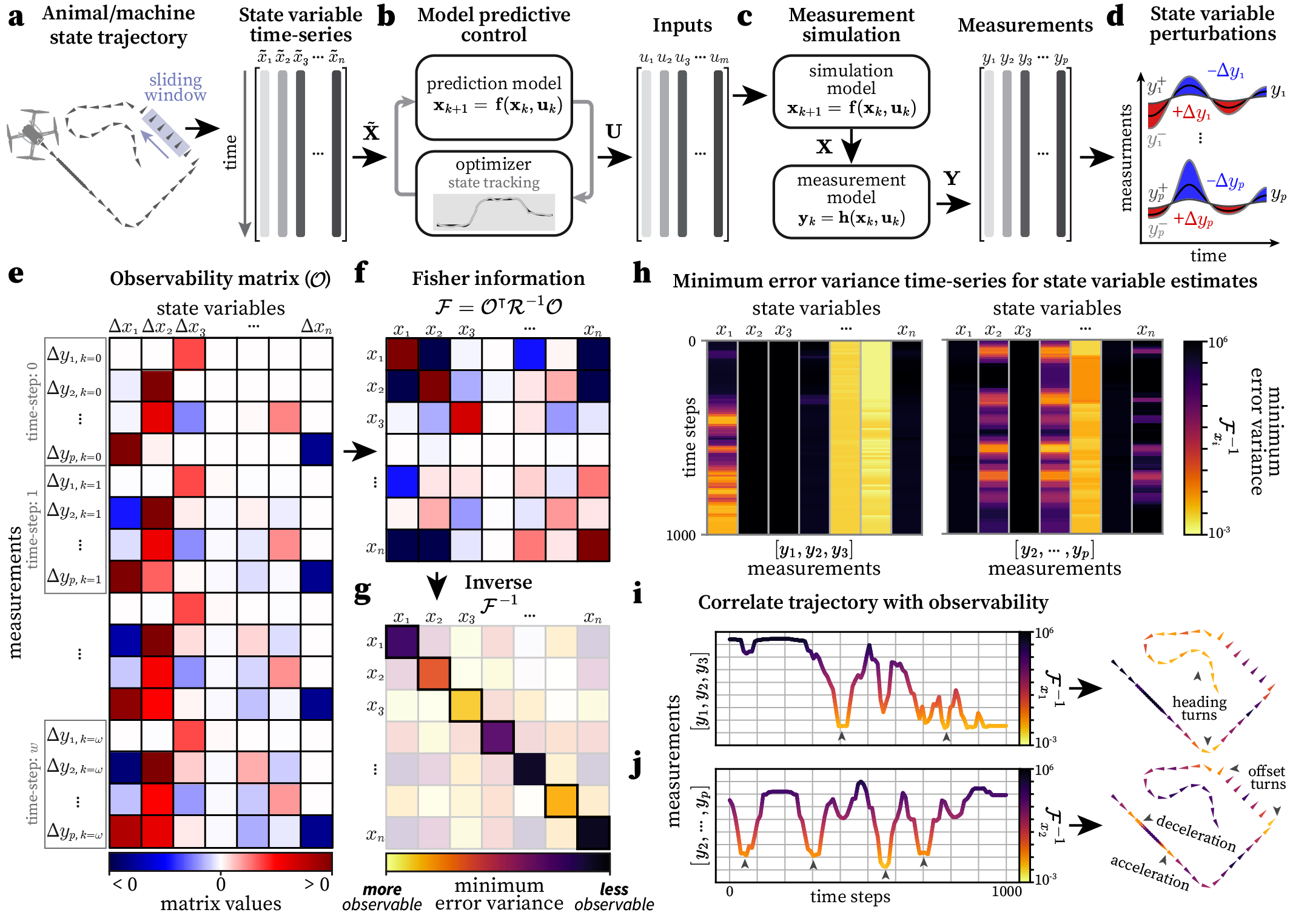}}
    \caption{\textbf{Overview of BOUNDS, a method for discovering active sensing motifs with empirical observability.} See Methods for a detailed step-by-step description. 
    \textbf{a.} A simulated (or measured) state trajectory (in this case, of a flying agent) that is described by a collection of time-series of observed state variables $\vect{\tilde{X}}$.
    \textbf{b.} Model predictive control (MPC) is used to find the time-series of control inputs $\vect{U}$ required to reconstruct the state trajectory in \textit{a} through a simulation model. 
    \textbf{c.} The MPC control inputs $\vect{U}$ are used to simulate the dynamics associated with the state trajectory, yielding the nominal reconstructed trajectory $\vect{X}$. A measurement model $\vect{h}(\vect{x}_k, \vect{u}_k)$ simulates the sensor measurements.
    \textbf{d.} Each initial state variable $x_{i,0}$ is perturbed in the positive $x_{i,0}^{i+}$ and negative $x_{i,0}^{i-}$ directions by $\epsilon$. Then the corresponding measurements over time, $\vect{Y}^{i+}$ and  $\vect{Y}^{i-}$, are computed. 
    The difference between these measurements is used to calculate a numerical Jacobian: $\Delta \vect{Y}^{i} / 2\epsilon$, yielding one column of the observability matrix $\mathcal{O}$.
    \textbf{e.} An illustration of an observability matrix $\mathcal{O}$, where the color grid indicates the values in the matrix (corresponding to \textit{d}). The columns indicate the difference of the initial state perturbation $\Delta x_{i,0}$ and the rows indicate the corresponding difference in measurements $\Delta \vect{y}_k$ over time for a discrete number of time steps $\timewindow$.
    \textbf{f.} An illustration of the Fisher information matrix, calculated as $\mathcal{F} = \mathcal{O}^{\intercal} \mathcal{R}^{-1} \mathcal{O}$, where $\mathcal{R}$ is the block diagonal sensor noise covariance matrix. Colors indicate matrix values as in \textit{e}.
    \textbf{g.} The inverse of the Fisher information matrix $\mathcal{F}^{-1}$, which encodes the minimum error covariance of the estimate of the state's initial condition $\vect{x}_0$. The diagonal elements represent the minimum error variance of each individual state variable---inversely correlated with the observability level---where smaller values indicate a higher observability level.
    \textbf{h.} The minimum error variance time-series (normalized units) of all state variables (diagonal of  $\mathcal{F}^{-1}$) over time for the first three sensors $[y_1, y_2, y_3]$ (left) and sensors $[y_2, \cdots,  y_p]$ (right). Note that some states are persistently observable or unobservable, while other states exhibit time-varying observability, which is dependent on the sensor set.
    \textbf{i.} The minimum error variance time-series of the first state variable $x_1$ for the first three sensors $[y_1, y_2, y_3]$ (left) and the corresponding correlation with the active sensing motifs in the simulated trajectory (right) from a. Time-series corresponds to the first column in \textit{h}. (left). By correlating the time-series with the trajectory, we reveal that only turns in heading improve the observability of $x_1$. To make the relationship between temporal patterns in observability and the features of the trajectory more intuitive, we shift the time series back in time by half of the sliding window length, $\timewindow/2$.
    \textbf{j.} Same as \textit{i} but for state variable $x_2$ and sensors $[y_2, \cdots,  y_p]$. Corresponds to the second column in \textit{h} (right). This sensor set renders acceleration/deceleration and turns offset to heading observable, but headings turns have no effect.
    }
    \createsubpanels{figure_1}{10}
    \label{fig:figure_1}
\end{figure*}

\section*{Results}

\subsection*{BOUNDS: Bounding Observability for Uncertain Nonlinear Dynamic Systems}

To build a foundation for methodically exploring how movement can enhance estimation of individual state variables, we developed BOUNDS. 
BOUNDS begins with two user inputs: a discrete time-series state trajectory $\vect{\tilde{X}}$---derived either from simulations or real data---describing agent or sensor movement and environmental variables (\autoref{fig:figure_1_a}), and a model defining the dynamics of the agent's movement ($\mathbf{x_{k+1}}=\mathbf{f}(\mathbf{x_k},\mathbf{u_k})$) and measurements ($\mathbf{y}=\mathbf{h}(\mathbf{x_k},\mathbf{u_k})$) (\autoref{fig:figure_1_b}). Our empirical approach accommodates any forward-simulable model---closed-form, physics engine, or data-driven. Control inputs can be provided, or in their absence, we apply model predictive control (MPC) to determine the inputs $\vect{U}$ that produce the nominal trajectory $\vect{X}$, from which we calculate the measurements $\vect{Y}$ with the model (\autoref{fig:figure_1_c}). 

To quantify how much information the measurements provide about each state variable, we determine the sensitivity of the measurements to small perturbations of each state (\autoref{fig:figure_1_d}). This is accomplished empirically by perturbing each state variable $x_{i,0}$ (one at a time) in both directions by a small amount $\epsilon$, without changing the control inputs from the nominal values, and recording how the subsequent measurements $\vect{Y}$ evolve \citep{Singh2005, Krener2009}. Repeating this process for each state variable yields the Jacobian $\Delta \vect{Y} / \Delta \vect{x}_0$, which constitutes the empirically determined observability matrix $\mathcal{O}$ \citep{Georges2020} (\autoref{fig:figure_1_e}). 

We then compute the Fisher information matrix $\mathcal{F} = \mathcal{O}^{\intercal} \mathcal{R}^{-1} \mathcal{O}$ (assuming no process noise)\citep{Boyacoglu2024}, where $\mathcal{R}$ accounts for measurement noise covariance (\autoref{fig:figure_1_f}). The Cramér–Rao bound states that $\mathcal{F}^{-1}$ provides the lower bound on estimation error variance, but for partially observable systems $\mathcal{F}$ is singular. A key innovation is our regularized inverse: we compute $\mathcal{F}^{-1} = \lim_{\lambda\to 0^+} [ \mathcal{F} + \lambda I_{n \times n}]^{-1}$. This ``Chernoff inverse'' \citep{chernoff1953locally} yields infinity for unobservable states while providing bounded error estimates for observable ones, unlike the traditional pseudo-inverse \citep{Penrose1955} that produces misleading values. In practice, we do not evaluate the limit, instead choosing $\lambda$ to be smaller than the largest non-zero singular value of $\mathcal{F}$. The diagonal elements of $\mathcal{F}^{-1}$ thus quantify minimum error variance for each state variable with meaningful physical units (\autoref{fig:figure_1_g}).

Calculating $\mathcal{F}^{-1}$ in short sliding windows of length $\timewindow$ reveals how observability evolves along trajectories (\autoref{fig:figure_1_h}--\autoref{fig:figure_1_j}). (Note that the time window need not be contiguous, see Methods.) In our toy example, changes in heading angle produce order-of-magnitude decreases in minimum error variance for state $x_1$ when using sensors $[y_1, y_2, y_3]$ (\autoref{fig:figure_1_i}, \autoref{fig:figure_1_h}). Conversely, state $x_2$ becomes observable during acceleration and offset turns, but requires sensors $y_2$ to $y_p$ and does not need $y_1$ (\autoref{fig:figure_1_j}, \autoref{fig:figure_1_h}). This illustrates a fundamental insight: each state variable can have distinct observability properties and active sensing motifs. BOUNDS thus reveals which combinations of sensor motion, available measurements, and state variables enable effective estimation.

\subsection*{Discovering active sensing motifs in a flying agent}

\begin{figure}[!th]
    \centerline{\includegraphics{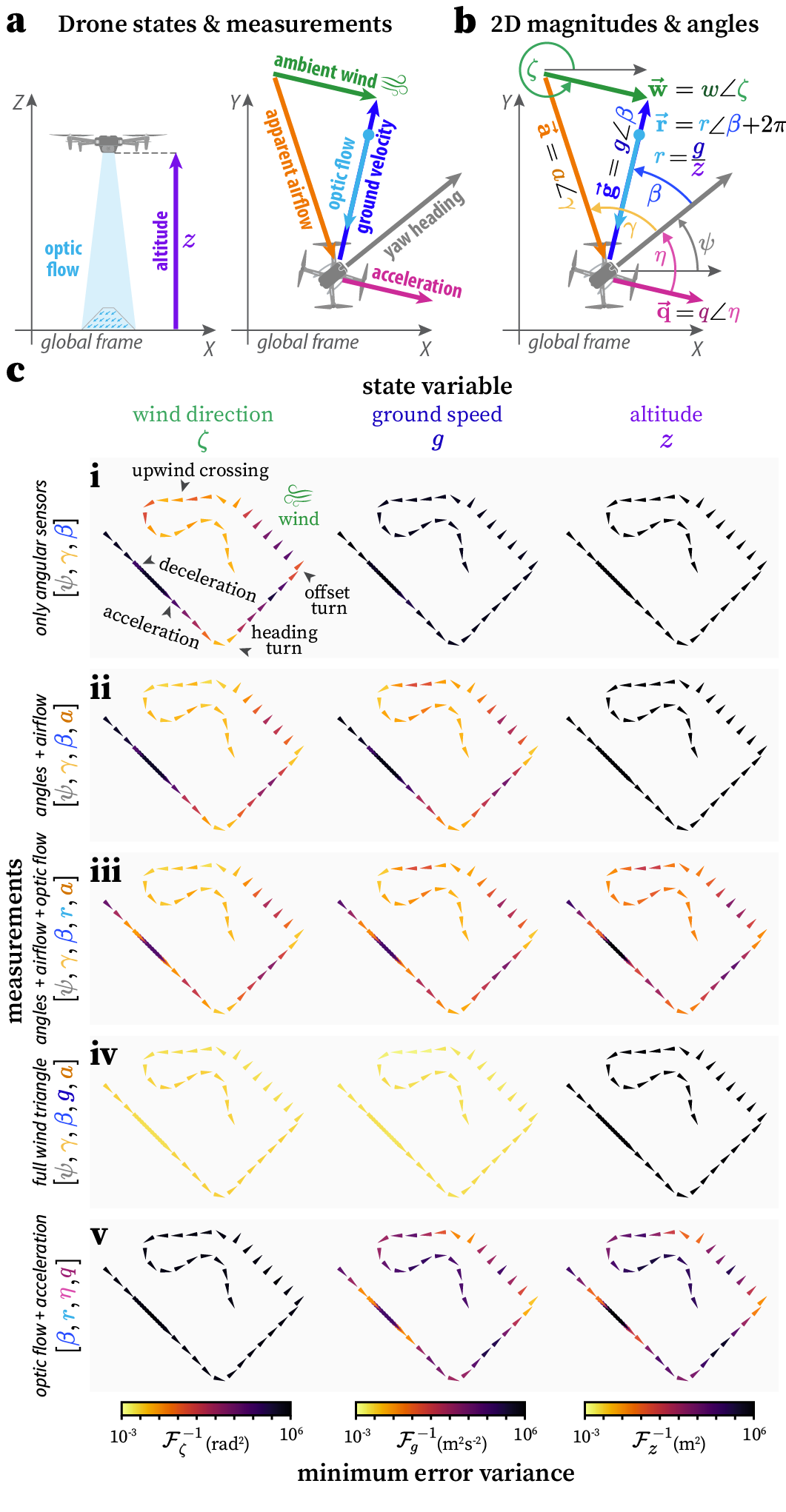}}
    \caption{\textbf{Evaluating putative active sensing motifs.}
    \textbf{a.} Illustration of the trigonometric and sensor kinematics of a flying agent in the presence of ambient wind in the XY plane (left) and XZ plane (right).
    \textbf{b.} Each vector quantity in \textit{a} is decomposed into a magnitude and angle, which can be considered separate measurements for observability analysis.
    \textbf{c.} The observability (minimum error variance) of the same simulated flight trajectory for different state variables (columns) and different subsets of measurements from \textit{b} (rows). Observability was calculated for \SI{0.5}{\second} time windows, corresponding to 5 discrete measurements ($\timewindow=5$). Note that there are complex interactions between the state variable of interest, the available measurements, and the movement patterns corresponding to an increase in observability.
    }
    \createsubpanels{figure_2}{3}
    \label{fig:figure_2}
\end{figure}

We applied BOUNDS to a 3D kinematic quadcopter model flying in ambient wind to discover active sensing motifs (see \hyperref[sec:Methods]{Methods}). The primary state variables consist of the agent's attitude (roll $\phi$, pitch $\theta$, yaw $\psi$), translational velocity ($v_x$, $v_y$, $v_z$) in the body-level frame, altitude ($z$), and ambient wind speed ($w$) and direction ($\zeta$). We assume that the agent has several body-level sensors: heading $\psi$, an air flow sensor to measure apparent air flow speed ($a$) and direction ($\gamma$), an accelerometer that provides the acceleration magnitude ($q$) and direction ($\eta$), and a vision system that measures ventral optic flow magnitude ($r=g/z$) and direction ($\beta$). These measurements make up components of the wind vector ``triangle" (\autoref{fig:figure_2_a}--\autoref{fig:figure_2_b}) experienced by any flying agent \citep{VanBreugel2021a}, and pose interesting active sensing questions that can be answered via BOUNDS.

We examined four movement motifs for our agent flying in constant wind: 1) constant course acceleration and deceleration, 2) turning while changing heading, 3) turning without changing heading (offset turn), and 4) a small turn through the upwind  direction (\autoref{fig:figure_2_c}). We then asked which states became more or less observable for these motifs depending on which sensors are available. For brevity, we focus on the state variables wind direction $\zeta$, altitude $z$, and ground speed $g$. Note that $g$ is not in the original state space of our model (see \hyperref[sec:Methods]{Methods}) and requires a coordinate transformation, which can be done post-hoc by applying the chain rule to transform $\mathcal{O}$ through the Jacobian of the coordinate change (see \hyperref[sec:Methods]{Methods}). A summary of the conditions for observability is given in Table \ref{tab:sensor_sets}, and discussed in the following paragraphs. Our results are not particularly sensitive to the model type: adding the complexity of a full dynamic quadcopter model yielded similar results (\autoref{fig:figure_supp_dynamic}). 

With only angular measurements (heading $\psi$, course direction $\beta$, apparent airflow angle $\gamma$), wind direction alone is observable---but \textit{only during heading changes}, not straight flight (\autoref{fig:figure_2_c}\romannumeral 1). This result is consistent with both analytical observability results \citep{VanBreugel2021a} and conclusions from optimal estimator designs \citep{VanBreugel2022}. Larger turns generally increased observability, though small turns through the upwind direction sufficed (\autoref{fig:figure_2_c}\romannumeral 1). Acceleration/deceleration and offset turn motifs did not markedly increase observability for this sensor set. 

Adding apparent airflow magnitude ($a$) made offset turns sufficient for wind direction and enabled ground speed estimation for many motifs (\autoref{fig:figure_2_c}\romannumeral 2). Adding optic flow magnitude ($r$) as well made acceleration and deceleration motifs viable for estimating both wind direction and ground speed, and rendered altitude observable across all motifs (\autoref{fig:figure_2_c}\romannumeral 3). Replacing the optic flow measurement with a direct ground speed measurement ($g$) rendered the wind direction and ground speed observable without any motif (even during straight flight) (\autoref{fig:figure_2_c}\romannumeral 4). 

Finally, we imitted the airflow sensor and examined a set consisting of optic flow magnitude ($r$) and direction ($\beta$) and acceleration magnitude ($q$) and direction ($\eta$), which proved suitable for estimating ground speed and altitude during acceleration/deceleration and offset turns (\autoref{fig:figure_2_c}\romannumeral 5). However, wind direction was never observable, and heading turns had a minimal effect on the observability (\autoref{fig:figure_2_c}\romannumeral 5).

\begin{table}[]
\caption{Observable state variables for different sensor sets. For each sensor set from \autoref{fig:figure_2_c}, we list the motion primitives (motifs) that can each independently render wind direction ($\zeta$), ground speed ($g$), and altitude ($z$) observable.}
\label{tab:sensor_sets}
\scriptsize
\setlength{\tabcolsep}{3pt}
\renewcommand{\arraystretch}{1.1}
\begin{tabular}{@{}lll@{\hspace{2pt}}>{\centering\arraybackslash}p{0.5cm}@{\hspace{0pt}}>{\centering\arraybackslash}p{0.5cm}@{\hspace{0pt}}>{\centering\arraybackslash}p{0.5cm}@{}}
\toprule
\textbf{Sensor Set} & \textbf{Symbols} & \textbf{Motif} & \multicolumn{3}{c}{\textbf{Obs. states}} \\
\cmidrule(l){4-6}
 & & & $\zeta$ & $g$ & $z$ \\
\midrule
angles only & $[\textcolor[HTML]{8B7D8B}{\psi}, \textcolor[HTML]{4169E1}{\beta}, \textcolor[HTML]{FFA500}{\gamma}]$ & heading change & $\checkmark$ & & \\
\addlinespace
add airflow & $[\textcolor[HTML]{8B7D8B}{\psi}, \textcolor[HTML]{4169E1}{\beta}, \textcolor[HTML]{FFA500}{\gamma}, \textcolor[HTML]{FF7F00}{a}]$ & heading change, offset turn & $\checkmark$ & $\checkmark$ & \\
\addlinespace
add optic flow & $[\textcolor[HTML]{8B7D8B}{\psi}, \textcolor[HTML]{4169E1}{\beta}, \textcolor[HTML]{FFA500}{\gamma}, \textcolor[HTML]{FF7F00}{a}, \textcolor[HTML]{00BFFF}{r}]$ & accel, heading change, offset turn & $\checkmark$ & $\checkmark$ & $\checkmark$ \\
\addlinespace
optic flow, accel. & $[\textcolor[HTML]{4169E1}{\beta}, \textcolor[HTML]{00BFFF}{r}, \textcolor[HTML]{FF69B4}{\eta}, \textcolor[HTML]{9B4B9B}{q}]$ & accel, offset turn & & $\checkmark$ & $\checkmark$ \\
\bottomrule
\end{tabular}
\end{table}

Although we use a flying agent to provide a concrete example, our results highlight a general principle applicable to many nonlinear partially observabile systems: observability is state-variable-specific and sensor-configuration-dependent. Unlike classical observability analyses designed to reveal the \textit{combinations} of states that are most observable (by analyzing the eigenvectors of the observability Gramian $\mathcal{O}^T\mathcal{O}$), BOUNDS reveals how optimal active sensing strategies can be uniquely tailored to target specific state variables. The motifs BOUNDS reveals can subsequently be used either to generate hypotheses about the role motifs may serve in organisms, or to inform control/estimation strategies and algorithms for autonomous agents with limited sensors. The same approach we show in our case study can be applied in a similar to any arbitrary system to discover measurement- and movement-dependent sensing motifs.

\begin{figure*}[!t]
    \centerline{\includegraphics{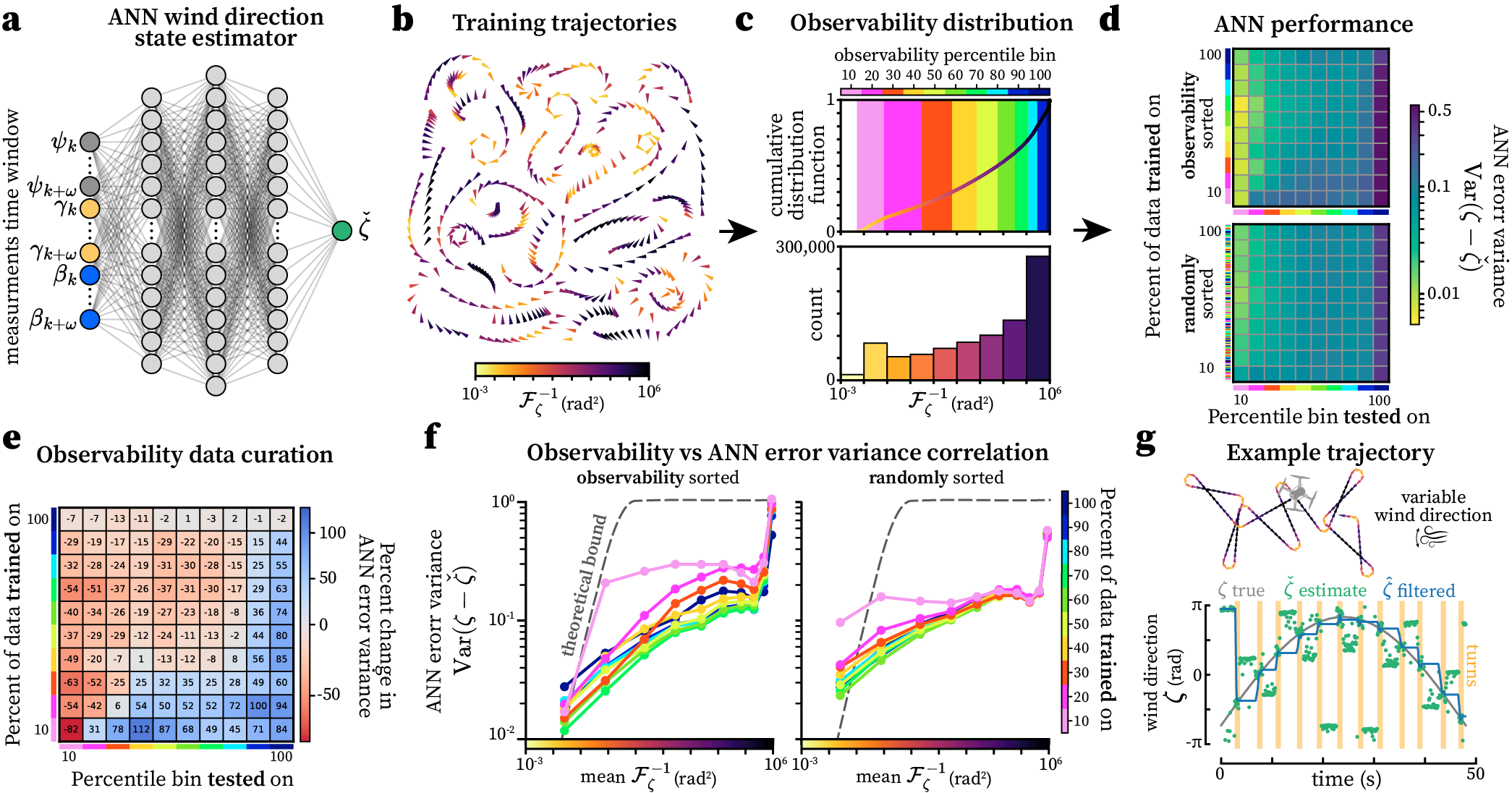}}
    \caption{\textbf{Observability-informed state estimation.}
    \textbf{a.}   Artificial neural network (ANN) state estimator for wind direction ($\mathcal{H}_{\zeta}$). Inputs are the yaw heading angle ($\psi$), apparent wind angle ($\gamma$), and course direction angle ($\beta$) in the time window $\timewindow$. 
    \textbf{b.} A subset of simulated flying agent trajectories used for training and testing the ANN state estimator. $N=40,000$ trajectories total.
    \textbf{c.} The distribution (bottom) and cumulative distribution function (top) of wind direction observability ($\mathcal{F}^{-1}_{\zeta}$) across all training and testing trajectories.
    \textbf{d.} A comparison of the testing error variance for the wind direction state estimator ANN ($\mathcal{H}_{\zeta}$) when trained on different percentages of the training data---for data sorted by observability (sorted high-to-low, top) and randomly sorted (bottom). The y-axis indicates how much data was used in training, and the x-axis indicates what bin percentile was used to test the ANN (testing data always sorted by observability).
    \textbf{e.} Same as \textit{d.} but showing the relative change in error variance between observability-sorted and randomly sorted training data. Red values ($<$ 0) indicate better performance when sorting by observability.
    \textbf{f.} The mean observability $\text{mean}(\mathcal{F}^{-1}_{\zeta})$ of each bin in the training dataset vs the ANN error variance $\text{Var}(\check{\zeta} - \zeta)$ for ANNs trained on different percentages of the training dataset for observability-sorted (left) and randomly sorted (right) training data. The theoretical bound of the ANN performance is shown as a dashed line (bounded at 1 for circular data; see Methods). 
    \textbf{g.} (top) An example simulated trajectory of a flying agent in the presence of time-varying wind direction. Color along trajectory indicates observability. (bottom) A comparison of true wind direction (${\zeta}$), ANN wind direction estimate ($\check{\zeta}$), and the observability-filtered estimate ($\hat{\zeta}$). Measurement noise covariance was set to $\mathcal{R} = 10^{-3} I_{p \times p}$.
    }
    \createsubpanels{figure_3}{7}
    \label{fig:figure_3}
\end{figure*}

\subsection*{Observability-informed state estimation}

How do we exploit the sporadic information provided by active sensing maneuvers? Here we develop a simple framework that leverages knowledge of the observability to determine when state estimates are likely to be accurate, and when they are not, and use this knowledge to generate an observability filtered estimate. Building such a filter involves four steps: constructing a state estimator, curating data by observability, training an observability estimator, and observability-informed filtering. We show an example application of each step for wind direction estimation.


\subsubsection*{State estimator}
We define a state estimator $\mathcal{H}_i$ for any state variable ${x}_{i}$
\begin{equation} \label{equ:state_estimator}
    \begin{aligned}
        \check{x}_{i,k} &= \mathcal{H}_{i}(\vect{y}_k, \dots, \vect{y}_{k-\timewindow},\vect{u}_k, \dots, \vect{u}_{k-\timewindow}, \mathcal{R}_{\timewindow_k})& \\ &=
        \mathcal{H}_{i}(\vect{Y}_{\timewindow_k},\vect{U}_{\timewindow_k}, \mathcal{R}_{\timewindow_k}).&
    \end{aligned}
\end{equation}
$\mathcal{H}_i$ is a function that maps the time-history of measurements $\vect{Y}_{\timewindow_k}$ (and optionally inputs $\vect{U}_{\timewindow_k}$ and measurement noise covariance $\mathcal{R}_{\timewindow_k}$) in the time window $\timewindow$ to the current state estimate $\check{x}_{i,k}$ at time step $k$. Notably, $\mathcal{H}_{i}$ can consider the entire time window $\timewindow$ at once, and is not sensitive to initialization, thus can potentially avoid the pitfalls associated with Kalman filtering. In principle, $\mathcal{H}_i$ could be any function, but we employ a feed-forward artificial neural network (ANN) for its generality, flexibility, and ability to operate without relying on a predefined dynamical model. We designed an ANN for estimating wind direction using angular measurements $[\psi, \gamma, \beta]$ (\autoref{fig:figure_3_a}, \autoref{fig:figure_supp_ANN_noise_a}), which yields accurate estimates \autoref{fig:figure_supp_ANN_noise_b}, but only during heading changes as predicted by our observability analysis (\autoref{fig:figure_2_c}\romannumeral 1).

\subsubsection*{Curating training data by observability}

The distribution of observability in the data used to train the ANN is critical: training on primarily unobservable data will yield poor results. Knowledge of active sensing motifs discovered by BOUNDS can help us selectively generate high quality training data, or curate a subset of an existing dataset, to efficiently train an accurate estimator. 

To train our wind direction estimator, we randomly simulated a diversity of trajectories spanning a large envelope of the state space and with a wide distribution of observability values (\autoref{fig:figure_3_b}--\autoref{fig:figure_3_c}, see \hyperref[sec:Methods]{Methods}). We separated the data into ten equally sized bins, sorted by the observability of wind direction (\autoref{fig:figure_3_c}). This allowed us to train ANNs on inputs with distinct distributions of observability. We analyzed the ANN wind direction estimate error variance $\text{Var}(\zeta - \check{\zeta})$ when trained on different percentages of the training dataset---once when the training data was sorted by observability level, and once when the data was randomly sorted (\autoref{fig:figure_3_d}, \autoref{fig:figure_supp_ANN_percentile}). 

Sorting data by observability allowed the curation of a smaller, but richer, dataset that increased ANN performance. Training on between the 40\% and 70\% most observable data led to the best overall performance (\autoref{fig:figure_3_e}). Furthermore, observability was a good predictor of ANN performance and correlated with the ANN error variance (\autoref{fig:figure_3_f}), though our ANN outperformed the theoretical bound. This is likely a result of the ANN learning biased heuristics from the training data. For instance, when the wind speed is much greater than the ground speed in our flying agent, the wind direction can be approximated as $\zeta \approx \psi + \gamma$, even during unobservable straight flight.

\subsubsection*{Observability estimator}

Knowing when $\mathcal{H}_i$ produces accurate estimates requires knowing the observability of ${x}_{i}$. However, calculating the true observability $\mathcal{F}^{-1}_{x_i}$ relies on knowledge of the full state $\vect{x}_{k}$ in advance. The agent only has access to $\vect{Y}_{\timewindow_k}$, $\vect{U}_{\timewindow_k}$, and  $\mathcal{R}_{\timewindow_k}$, therefore $\vect{x}_{k}$ is not directly available. We take a data-driven approach to solve this problem by constructing an observability estimator $\mathcal{G}_i$ of the form
\begin{equation} \label{equ:observability_estimator}
   \check{ \mathcal{F}}^{-1}_{x_{i, k}}  = \mathcal{G}_i(\vect{Y}_{\timewindow_k}, \vect{U}_{\timewindow_k}, \mathcal{R}_{\timewindow_k}),
\end{equation}
where $\check{\mathcal{F}}^{-1}_{x_{i, k}}$ is an estimate of the current observability of  ${x}_{i}$ at time step $k$. We can similarly use an ANN to construct $\mathcal{G}_i$---however, for simple active sensing motifs there may be a heuristic relationship between a measured movement, e.g. turning, and observability that can be exploited to develop a less complex model. 

We trained a wind direction observability estimator ANN (\autoref{fig:figure_supp_ANN_noise_c}) using the entire distribution of our training data. At low noise, the estimator predicted true observability within an order of magnitude, capturing turn-associated spikes (\autoref{fig:figure_supp_ANN_noise_d}); higher noise caused overestimates because it is difficult to tease apart ``good" changes in measurements due to active sensing and ``bad" changes due to noise.

\subsubsection*{Observability-informed filtering}
The final step leverages observability estimates $\check{ \mathcal{F}}^{-1}_{x_i}$ to adaptively filter raw state estimates $\check{x}_{i,k}$. The filtered state estimate $\hat{x}_{i,k}$ should not change quickly when observability is low and $\check{x}_{i,k}$ is inaccurate, but should quickly converge to $\check{x}_{i,k}$ when observability is high and $\check{x}_{i,k}$ is accurate. A simple solution is an adaptive low-pass filter with coefficients dynamically scaled by the estimated observability,
\begin{equation} \label{equ:observability_filter}
        \hat{x}_{i,k} = \alpha_{k} \; \check{x}_{i,k} + (1 - \alpha_{k}) \; \hat{x}_{i,k-1}, 
\end{equation}
where $\hat{x}_{i,k}$ is the current filtered estimate of the state $x_i$, $\hat{x}_{i,k - 1}$ is the filtered estimate at the previous time step, and $\check{x}_{i,k}$ is the current raw estimate of the state from $\mathcal{H}_i$ (\autoref{fig:figure_supp_observability_filter}). The coefficient $\alpha_k$ is a time-varying weight between $0-1$ that determines how much to update the current filtered estimate with the current raw estimate. $\alpha_k$ can be determined from an inverted and normalized function of the estimated observability $\check{\mathcal{F}}^{-1}_{x_{i, k}}$.

We applied our observability filter to a trajectory with time-varying  wind direction, which the ANNs had not been trained on (\autoref{fig:figure_3_g}). This trajectory was designed to only have sporadic bouts of active sensing (turns) to make estimation more challenging. We also injected sensor noise into the measurements $\vect{Y}_{\timewindow_k}$. As expected, the ANN raw estimate ($\check{\zeta}$) was accurate only during turns; between turns, estimates were consistently offset (\autoref{fig:figure_3_g}). However, the filtered estimate ($\hat{\zeta}$) quickly converged to the true value of wind direction (${\zeta}$) during a turn (\autoref{fig:figure_3_g}). With sufficiently frequent turns the observability-filtered estimate could track the changing wind direction, although high enough measurement noise did lead to worse performance (\autoref{fig:figure_supp_ANN_noise_e}--\autoref{fig:figure_supp_ANN_noise_g}).

\subsection*{Augmented Information Kalman Filter}

\begin{figure*}[p!]
    \centerline{\includegraphics{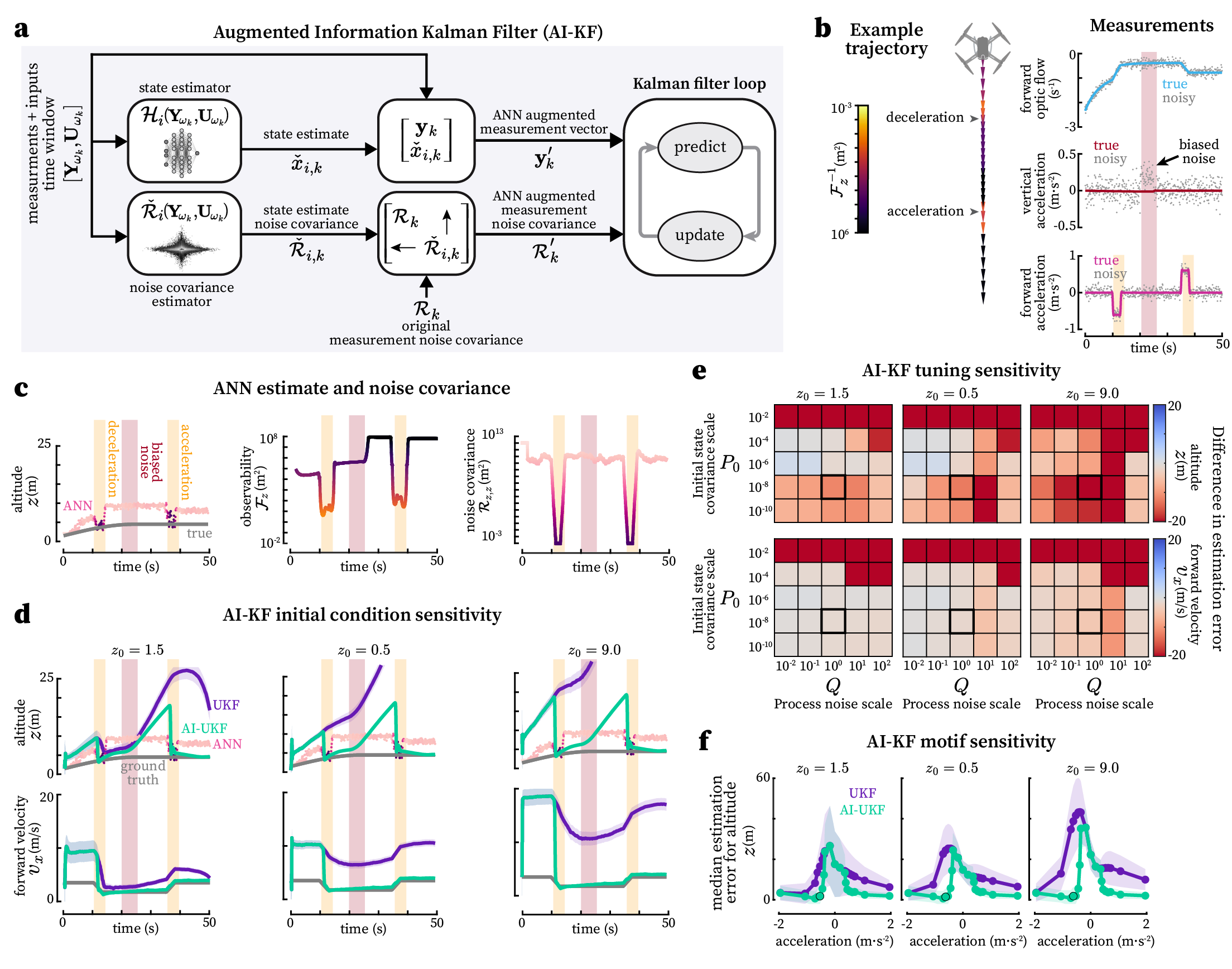}}
    \caption{\textbf{Augmented Information Kalman Filter.}
    \textbf{a.} Overview of the Augmented Information Kalman Filter (AI-KF) state estimation framework. The measurement vector $\vect{y}_k$ is augmented with a data-driven state estimate $\check{x}_{i,k}$. The corresponding state noise covariance is estimated $\check{\mathcal{R}}_{i, k}$ and used to dynamically adjust the full augmented noise covariance matrix $\mathcal{R}'_k$ based on the state estimate accuracy (or observability).
    \textbf{b.} (left) The AI-KF framework is applied to estimate altitude for a simulated trajectory consisting of straight flight with a brief deceleration and sequential acceleration. Altitude is observable during acceleration/deceleration (see \autoref{fig:figure_2_c}\romannumeral 5) given measurements of forward optic flow, forward acceleration, and vertical acceleration. 
    (right) Noisy measurements used in our simulation with additive zero-mean Gaussian noise with a variance of $10^{-2}$ for each measurement. There is also a brief period of biased (nonzero-mean) noise added to the vertical acceleration measurement.    
    \textbf{c.} Artificial neural network estimates of $\check{z}$, observability of $z$, and the noise variance of the augmented altitude measurement $\mathcal{R}_{z,z}=\mathrm{Var}(\tilde{\check{z}})$ for the trajectory shown in \textit{b}. $\mathcal{R}_{z,z}$ was determined according to equations \ref{eq:R_map}- \ref{eq:R_ratio}. 
    \textbf{d.} A comparison of an Unscented Kalman Filter (UKF, \textit{purple}) and Augmented Information Unscented Kalman Filter (AI-UKF, \textit{green}) state estimation results with varied initial altitude state estimates, for the trajectory in \textit{b}. The true value of $z_0$ was \SI{1.5}{\meter}. The shading around the curve indicates the 2-$\sigma$ estimate bound. 
    \textbf{e.} A comparison of the median error of the UKF and AI-UKF for varying process noise covariance matrix ($Q$) scales and varying initial state covariance matrix ($P_0$) scales for varying values of the initial altitude estimate. The results in \textit{d} correspond to the square with a bold outline.
    \textbf{f.} Comparison of the UKF and AI-UKF median error across different mean accelerations for varying values of the initial altitude estimate. The AI-UKF always performs better or equal to the vanilla UKF. The advantage of the AI-UKF is most pronounced for small but non-zero levels of acceleration.
    }
    \createsubpanels{figure_4}{6}
    \label{fig:figure_4}
\end{figure*}

The observability-filtering approach does not take full advantage of the dynamic relationships between states and inputs, and lacks the statistical rigor of model-based methods like the Kalman filter (KF). The observability-filter essentially acts like a zero-order hold during periods of low observability, but a KF can potentially make use of ``dead-reckoning" to track states with reasonable accuracy in these same periods \citep{Brossard2020}. Kalman filtering, however, comes with its own challenges for weakly observable and nonlinear systems \citep{Butcher2017, Bocquet2017}:  an incorrect guess for the initial state $\hat{\vect{x}}_{k=0}$  can drive the new estimate in the wrong direction. Both the predicted state and innovation terms in the KF update equation,

\begin{equation}  \label{eq:kalman}
\underbrace{\hat{\vect{x}}_{k+1}}_{\substack{\text{new state}}}
= 
\underbrace{\vect{f}(\hat{\vect{x}}_{k},\vect{u}_{k})}_{\substack{\text{predicted state}}}
+ \underbrace{K}_{\substack{\text{Kalman} \\ \text{gain} }}
(\underbrace{\tilde{\vect{y}}_{k} - \vect{h}(\hat{\vect{x}}_{k},\vect{u}_{k})}_{\substack{\text{innovation}}}),
\end{equation}
rely on an accurate $\hat{\vect{x}}_{k}$, so the KF will struggle to converge when $\hat{\vect{x}}_{k}$ is incorrect . Additionally, the initial state covariance $P_{k=0}$  influences the computation of the Kalman gain $K$, so a wrong guess for $P_{k=0}$ can also lead to unexpected behavior. For linear and observable systems this is generally not a problem, but when observability is low or sparse in time, and dynamics are nonlinear, a KF filter is far from an optimal estimator. Because only one measurement at a time is used to update the state estimate (\autoref{eq:kalman}), if $\hat{\vect{x}}_{k=0}$ or $P_{k}$ was wrong in the past, then potentially useful prior information is lost. Alternatively, an estimator that is able to consider the entire time-history of measurements---such as our $\mathcal{H}$ function---may be able to outperform a KF.

Our Augmented Information Kalman Filter (AI-KF) augments traditional Kalman filtering with data-driven state estimates from $\mathcal{H}$ (\autoref{fig:figure_4_a}). The AI-KF works by augmenting a KF with extra information from a data-driven state estimator, and then telling the KF when to trust this information more or less with a time-varying noise covariance matrix that is adjusted based on an observability estimate. Unlike heuristic ANN-KF hybrids \citep{Li2021,Choi2023,Ghosh2024,krishnan2015deep}, the AI-KF weights augmented measurements by their Cramér-Rao bound, grounding information fusion in observability theory. The main benefit is that the augmented information considers an arbitrarily long time-history of measurements that can make up for a ``bad" initialization of the KF. This observability-based weighting also enables interpretability, making \textit{when} and \textit{why} the AI-KF outperforms traditional filtering explicit.

First, the original Kalman filter measurement vector $\vect{y}$ is augmented with a state estimate $\check{x}_{i}$ from a data-driven state estimator $\mathcal{H}_i$ at each time step,
\begin{equation}  \label{eq:aikf_y}
\vect{y}^{\prime} = 
    \left[
        \begin{aligned}
            \vect{y} \\
            \check{x}_{i}
        \end{aligned}
    \right].
\end{equation}
In principle, the measurement vector can be augmented with any subset of data-driven state variable estimates ${\check{x}_{0}, \dots, \check{x}_{n}}$. However, for brevity, we show the process for a single state variable. Next, the original Kalman filter noise covariance $\mathcal{R}$ is augmented with the noise covariance for the data-driven state estimate $\check{x}_{i}$
\begin{align*} \label{eq:aikf_R}
 \mathcal{R}^{\prime}=&\begin{bmatrix}
  \mathcal{R} & \operatorname{Cov}(\mathbf{y},\tilde{\check{x}}_i) \\[4pt]
  \operatorname{Cov}(\mathbf{y},\tilde{\check{x}}_i)^\top & \operatorname{Var}(\tilde{\check{x}}_i)
\end{bmatrix},
\end{align*}
where $\tilde{\check{x}}_i$ is the data-driven estimation error for the state variable $x_i$. The augmented portion of $\mathcal{R}^{\prime}$, which we will refer to as $\check{\mathcal{R}}_i$, encodes the covariance of $\check{x}_{i}$ both with itself $\big(\operatorname{Var}(\tilde{\check{x}}_i)\big)$ and all other measurements $\big(\operatorname{Cov}(\mathbf{y},\tilde{\check{x}}_i)\big)$. 

The critical step is constructing $\check{\mathcal{R}}_i$ to vary in time based on the accuracy of the state estimate $\check{x}_i$. This task can be accomplished multiple ways. If $\mathcal{H}_i$ is unbiased and \textit{efficient}\footnote{An \textit{efficient }(unbiased) estimator is an estimator for which the error variance reaches the Cramér–Rao Bound\citep{crassidis2004}.}, then the augmented portion of $\mathcal{R}^{\prime}$ is equivalent to the observability from $\mathcal{F}^{-1}$, which can be estimated with $\mathcal{G}_i$. However, it is more statistically consistent to construct a covariance function by performing a post hoc analysis of the state estimator $\mathcal{H}_i$ itself to determine the error covariance as a function of the estimator inputs $\check{\mathcal{R}}_i(Y_{\timewindow}, U_{\timewindow})$. Notably, the off-diagonal elements of $\check{\mathcal{R}}_i$ will generally be non-zero because $\check{x}_{i}$ is a function of $\vect{y}_{k}$ (in addition to prior measurements), but the diagonal element $\check{\mathcal{R}}_{i,i}$ will have the most impact. Because the Kalman filter assumes that each measurement (including $\check{x}_{i}$ is independent of the prior state estimate (including $\hat{x}_{i-1}$), there is a potential for double counting information once the KF has converged to an accurate state estimate. Although this will not harm the accuracy of the state estimate, it can artificially decrease the error variance. In the methods we describe a simple resolution to this that allows the AI-KF to converge to a classic KF when the two are in agreement. 

\subsubsection*{Applying AI-KF to estimate altitude}
We implemented our AI-KF framework for a flying agent to estimate altitude $z$ and forward velocity $v_x$ from only forward ventral optic flow $r_x$, forward acceleration $\dot{v}_x$, and vertical acceleration $\dot{v}_z$ (see \hyperref[sec:Methods]{Methods}). Observability analysis tells us that this estimation problem is solvable given non-zero acceleration (\autoref{fig:figure_2_c}\romannumeral 5) \citep{VanBreugel2014a}. It is a more challenging problem than wind direction estimation (see difference in observability in \autoref{fig:figure_2_c}) and more clearly illustrates the benefit of the AI-KF. In this example, we augment an Unscented Kalman filter (UKF), which is appropriate for nonlinear dynamics---although our method is applicable to most Kalman filter extensions. 

We used a simplified 2D kinematic model for the AI-UKF with only the state variables altitude ($z$), vertical velocity ($v_z$), and forward velocity ($v_x$)
\begin{equation} \label{equ:model_2d}
\begin{aligned}
    \vect{\dot{x}} =
        \begin{bmatrix} 
            \dot{z} \\
            \dot{v}_z  \\
            \dot{v}_x  \\
        \end{bmatrix}
        =
        \begin{bmatrix}
            v_z \\
            u_z \\
            u_x \\
        \end{bmatrix},
    \end{aligned}
\end{equation}
with the measurement model
\begin{equation} \label{equ:model_2d_measurements}
\begin{aligned}
    \vect{y} =
        \begin{bmatrix} 
            r_x \\
        \end{bmatrix}
        =
        \begin{bmatrix}
            -v_x/z \\
        \end{bmatrix},
    \end{aligned}
\end{equation}
where the inputs $u_z$ and $u_x$ are measured accelerations. We chose this model for its simplicity and generality---it applies to virtually any flying agent.

We trained a data-driven ANN estimator that accurately estimates altitude from two-second windows of optic flow and forward acceleration (\autoref{fig:figure_supp_altitude_ANN_a}--\autoref{fig:figure_supp_altitude_ANN_b}), and augmented the KF measurement vector with this altitude estimate,
\begin{equation}  \label{eq:aikf_altitude}
\vect{y}^{\prime} = 
    \begin{bmatrix}
            r_x \\
            \check{z}
    \end{bmatrix}.
\end{equation}
Guided by insight from our observability analysis (\autoref{fig:figure_supp_altitude_ANN_c}--\autoref{fig:figure_supp_altitude_ANN_d}), 
we developed an empirical relationship between the altitude estimate's noise variance $\check{\mathcal{R}}_{z, z}$ and the mean forward acceleration magnitude from the ANN time window (\autoref{fig:figure_supp_altitude_ANN_c}--\autoref{fig:figure_supp_altitude_ANN_d}). To match the rapid decline of altitude estimation errors with increasing horizontal accelerations we chose a function with a negative exponential form, 

\begin{equation}  \label{eq:R_map}
    \begin{aligned}
        \check{\mathcal{R}}_{z,z} &= {(\rho_{\min})}^{1-\sigma}\,{(\rho_{\max})}^{\sigma},
    \end{aligned}
\end{equation}
where $\rho_{\min}$ and $\rho_{\max}$ are tunable positive parameters that set the absolute minimum and maximum of $\check{\mathcal{R}}_{z, z}$. The exponent $\sigma$ is a normalized function of the lateral acceleration in the ANN time window given by
\begin{equation}  \label{eq:R_ratio}
    \begin{aligned}
        \sigma &= \frac{\max(|\dot{v}_{x}|)-\text{mean}(|\dot{v}_x|_{\timewindow})}{\max(|\dot{v}_{x}|)-\min(|\dot{v}_{x}|)}.
    \end{aligned}
\end{equation}
The value of $\sigma$ lies in the range $[0,1]$, ensuring that for windows with horizontal accelerations close to the maximum $\sigma=0$ and $\rho_{\min}$ will dominate, leading to a small value for $\check{\mathcal{R}}_{z, z}$, allowing the ANN altitude estimate to have a strong influence on the Kalman filter update equation. For our system, we chose $\rho_{\min}=10^{-3}$ to allow the accurate augmented altitude measurements have a large influence when acceleration was large, and we set $\rho_{\max}=10^{12}$ to ensure that periods of low acceleration did not affect the AI-UKF.  For simplicity, we set the off-diagonal elements of $\check{\mathcal{R}}_z$ to 0. 

To Test the AI-UKF, we simulated a straight trajectory with brief (\SI{3}{\sec}) deceleration and acceleration (\autoref{fig:figure_4_b}), zero-mean Gaussian noise throughout, and biased noise between maneuvers (\autoref{fig:figure_4_c}). As expected, the ANN estimates accurately predicted the correct altitude only during periods of relatively high observability (\autoref{fig:figure_4_c}).

\begin{figure}[!th]
    \centerline{\includegraphics{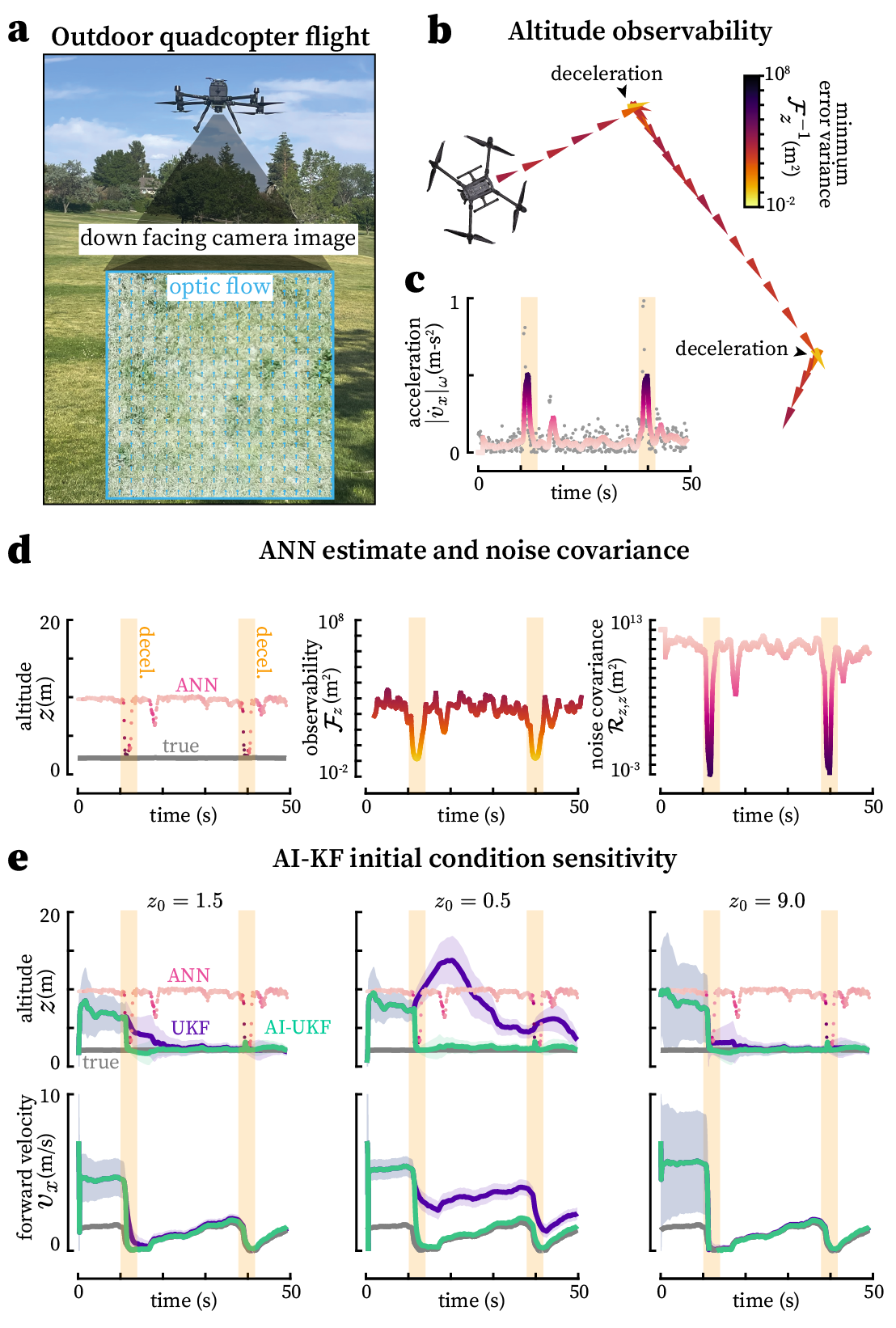}}
    \caption{\textbf{Augmented Information Kalman Filter application for outdoor quadcopter flight.}
    \textbf{a.} Quadcopter (DJI Matrice 300 RTK) flying in an outdoor environment with a representative ventral camera image and computed optic flow vectors.
    \textbf{b.} An experimental measured quadcopter trajectory with estimated altitude observability computed using the approach in \autoref{fig:figure_1}, illustrating how our BOUNDS framework can be used to determine observability for measured, rather than simulated, trajectories.
    \textbf{c.} Same as \autoref{fig:figure_4_c} but for experimental data from the trajectory in \textit{b}. As in \autoref{fig:figure_4_c}, the clear benefit of the AI-UKF is its robustness to different initial conditions and faster convergence to correct estimates. Ground truth velocity and altitude was provided by GPS measurements. 
    }
    \createsubpanels{figure_5}{9}
    \label{fig:figure_5}
\end{figure}

Applying both UKF and AI-UKF while varying initial altitude guesses (\autoref{fig:figure_4_d}) showed poor UKF performance overall. For some initial conditions, the UKF altitude estimate moved in the correct direction during the deceleration (first period of high observability) (\autoref{fig:figure_4_d}). However, the UKF struggled to converge to the true altitude after being thrown off by the biased noise  (\autoref{fig:figure_4_d}). In contrast, the AI-UKF was able to converge quickly and smoothly to the true altitude during both periods of high observability (\autoref{fig:figure_4_d}). Due to the coupling of altitude and forward velocity in the optic flow measurement ($r_x = v_x / z$), the forward velocity estimate also converged close to the true value for the AI-UKF, but again not for the  UKF (\autoref{fig:figure_4_d}). This illustrates how just a single data-driven estimator may be needed to estimate correlated state variables via the AI-KF.

In a sensitivity analysis, we examined how initial covariance $P_{k=0}$ and process noise $Q$ affected relative performance of the AI-UKF compared to the UKF. The AI-UKF consistently outperformed the UKF in terms of the median error after the first acceleration  (\autoref{fig:figure_4_e}). Although it may be possible to choose a specific combination of $\vect{x}_{k=0}$,  $P_{k=0}$, and $Q$ for a regular UKF that approximately matches the AI-UKF performance, there is no universal choice that will work on all trajectories and initializations (\autoref{fig:figure_4_e}). Therefore, the AI-UKF provides a degree of robustness to these initialization and hyperparameter choices.

The observability level---here, acceleration magnitude---drives the AI-UKF advantage. For large enough accelerations/decelerations, both the UKF and AI-UKF performed equally well (\autoref{fig:figure_4_f}). However, the UKF struggled with smaller accelerations/decelerations, while the AI-UKF was able to handle these instances without issue (\autoref{fig:figure_4_f}). For very small or zero accelerations the UKF and AI-UKF were equivalent because there was not enough observability for the augmented data-driven state estimate to have an effect  (\autoref{fig:figure_4_f}). Overall, the AI-UKF is most beneficial during periods of medium observability where a UKF might fail.

Fianlly, we validated the AI-UKF on real data. We collected optic flow and acceleration measurements from a quadcopter flying with approximately constant altitude and sporadic decelerations/accelerations, and then computed the observability of the trajectory via BOUNDS (\autoref{fig:figure_5_a}--\autoref{fig:figure_5_b}). During these decelerations (\autoref{fig:figure_5_c}) the quadcopter also changed course (\ref{fig:figure_5_b}). However, because our analysis is restricted to the velocity, optic flow, and acceleration along the direction of travel, these changes in course have no impact on the observability or estimates (\autoref{fig:figure_2_c}\romannumeral 5) . The same data-driven altitude estimator trained on simulated trajectories (\autoref{fig:figure_supp_altitude_ANN}) yielded impressive accuracy on real measurements during periods of high observability (\autoref{fig:figure_5_d}). However, when applying a regular UKF we observed similar results to our simulations, wherein the filtered estimates would sometimes not converge for certain initializations (\autoref{fig:figure_5_e}). In contrast, AI-UKF estimates converged quickly to the true altitude and forward velocity (\autoref{fig:figure_5_e}). For some initializations, the AI-UKF converged in seconds whereas the UKF did not converge within the $\sim$50-second trajectory (\autoref{fig:figure_5_c}). Overall, we suggest that in situations with limited or sparse sensor data, like in GPS-denied environments, estimating mission-critical state variables can benefit greatly from the AI-KF framework.

\section*{Discussion}

Decoding and exploiting active sensing requires tools that bridge formal observability theory with real-world complexity. BOUNDS builds this bridge by empirically quantifying the observability of individual state variables along trajectories for nonlinear, partially observable systems. Drawing on these observability insights, the Augmented Information Kalman filter demonstrates how to statistically merge classic and data-driven estimates through an observability-weighted function. This provides stability during periods of poor observability while exploiting information from sporadic sensing maneuvers. By accessing a larger history of information, this approach enables recovery from poor initialization that would corrupt sequential nonlinear filters. Beyond methodology, our work also reveals a key active sensing principle: state variables often have distinct observability properties requiring tailored strategies for sensor selection and movement. For example, for flying agents---be they machines or organisms---wind direction estimation is best supported by changes in course, whereas estimating altitude can be achieved solely through straight line acceleration (\autoref{fig:figure_2}). 

Our framework facilitates bidirectional insights for understanding natural navigation and advancing autonomous systems. In biology, a critical aspect of active sensing is knowing \textit{when} to perform it. Many animals appear to balance dedicated sensing maneuvers with other navigation goals while considering energetic cost \citep{chen2020tuning} through exploration-exploitation trade-offs triggered when estimation error reaches a threshold \citep{Biswas2023}. BOUNDS provides a principled framework for generating related predictions: by computing observability along behavioral trajectories, we can quantify when specific state estimates will degrade and which movements would restore accuracy. For example, our analysis predicts that wind direction estimation degrades during straight flight but is restored by heading changes---testable via closed-loop experiments where heading turns are blocked or disrupted before flies turn upwind \citep{Stupski2024}. Because neurons often multiplex sensory signals with complex filter dynamics \citep{chen2011representation}, it can be difficult to understand which navigational variables can be extracted. BOUNDS can help by revealing which combinations of movement and neural activity patterns are sufficient for estimating specific state variables, generating testable hypotheses about computational roles without requiring explicit decoding \citep{May2025compact}.

How animals integrate sporadic information from active sensing bouts into continuous estimates remains an open question. Our observability filter and AI-KF provide a theoretical structure for exploring this. If neural circuits implement observability-weighted integration similar to the AI-KF, then state estimates should update rapidly during high-observability movements but remain stable during low-observability periods---a prediction testable through neural recordings during naturalistic behavior. Already some evidence supports observability-gated updates: \textit{Drosophila} throttle learning plasticity in the central complex based on turning \citep{Fisher2022a}, a motif we find corresponds to high observability for many state variables (Fig. \ref{fig:figure_2_c}), suggesting that learning may be selectively enabled during observable epochs. 

Model-based estimation remains standard in robotics \citep{urrea2021kalman}, yet nonlinear tasks increasingly use windowed optimization \citep{VanBreugel2022,Shen2015,mourikis2007multi} or data-driven methods \citep{pellerito2024deep}---approaches that often sacrifice stability guarantees, simplicity, and interpretability \citep{jin2021new}. The AI-KF bridges these worlds through a statistically principled approach that is inherently modular: individual data-driven estimators (e.g., monocular SLAM, deep visual odometry) can be added or removed in the AI-KF structure without retraining. While not universally superior, the AI-KF excels for systems with sparse and dynamic observability that challenge traditional methods. Together, BOUNDS and the AI-KF enable engineers to develop machines that use active sensing to minimize required sensors, plan informative trajectories, and design robust estimators to approach the versatility observed in biological systems.

\newpage
\clearpage
\section*{Methods}
\label{sec:Methods}

\subsection*{BOUNDS step by step detail}

\subsubsection*{1. Determine state variables and collect state trajectory time-series}
Consider a discrete time-series state trajectory $\vect{\tilde{X}}$ consisting of a sequence of observed state vectors $\vect{\tilde{x}_k}$ that describe the movement of an agent or sensor, as well as environmental variables (\autoref{fig:figure_1_a}). The primary objective is to compute a new time-series that describes the observability of each individual variable $\tilde{x}_{i,k}$ along this trajectory, which can be used to evaluate the efficacy of putative active sensing motifs. All relevant state variables that describe the behavior and environment of the agent must be included, although the coordinates of the state variables can be optionally transformed in \textit{Step 6}. The state trajectory can be simulated for testing hypotheses, or measured from real data to evaluate the observability properties of an agent's underlying movement patterns.

\subsubsection*{2. Define a model}
Once a collection of state variables is determined, a model of the agent's dynamics must be defined. In general, a good model accounts for 1) the inputs that control locomotion (e.g, forces and torques) 2) basic physical properties such as mass, inertia, damping, etc., and 3) sensory dynamics that model what an agent is measuring (\autoref{fig:figure_1_b}). We describe our approach for a system with nonlinear deterministic dynamics, and stochastic measurements as
\begin{equation} \label{equ:system_dynamics}
\begin{aligned}
    \vect{x}_{k+1} &=\vect{f}(\vect{x}_k,\vect{u}_k), \\
    \vect{\tilde{y}}_k &= \vect{h}(\vect{x}_k, \vect{u}_k) + \vect{v}_k,
\end{aligned}
\end{equation}
where $\vect{x}_k$ is the true state vector representing $n$ state variables, $\vect{u}_k$ is the input vector, and $\vect{\tilde{y}}_k$ is the measurement vector corresponding to $p$ measurements at time step $k$. The function $\vect{f}(\vect{x}_k, \vect{u}_k)$ describes the system dynamics. The function $\vect{h}(\vect{x}_k,\vect{u}_k)$ describes the measurement dynamics, and $\vect{v}_k$ is zero-mean additive Gaussian noise with covariance matrix $R_k$. The effect of the noise on observability is taken care of in \textit{Step} 5. \textit{Steps} 3-4 only use the deterministic measurement equation $\vect{y}_k = \vect{h}(\vect{x}_k, \vect{u}_k)$. A benefit of our empirical approach is that the model does not have to be closed-form, and can instead be a simulation, computational, or data-driven model \citep{Singh2005, Krener2009}---for instance, a physics engine model of \textit{Drosophila} \citep{Vaxenburg2024} or finite element model of a flapping wing \citep{Boyacoglu2024a}. The only requirement is the ability to simulate the system forward in time. 

\subsubsection*{3. Reconstruct state trajectory through model predictive control}
If control inputs are not known, such as for an observed trajectory, we apply model predictive control (MPC) \citep{Camacho2007} to find the inputs $\vect{U}$ required to reconstruct the state trajectory with the chosen model (\autoref{fig:figure_1_b}). This step is critical for nonlinear or time-varying models where observability can change as a function of the state and input values. The MPC determined inputs are used to simulate the nominal trajectory $\vect{X}$ (which should be very similar to the observed trajectory $\vect{\tilde{X}}$) and associated measurements $\vect{Y}$ (\autoref{fig:figure_1_c}).

\subsubsection*{4. Construct observability matrix}
To quantify the observability of individual state variables, we next compute the observability matrix $\mathcal{O}$ empirically. $\mathcal{O}$ principally encodes information about the sensitivity of the sensor measurements  $\vect{y}_k$ to small changes in an initial state $\vect{x}_0$. It can be compactly represented in discrete-time as the Jacobian of the measurements over time with respect to the initial state, $\Delta \vect{Y}_{\timewindow} / \Delta \vect{x}_0$ \citep{Georges2020}. Here we use $\vect{Y}_{\timewindow}$ to denote the collection of $p$ measurements for each time step from $k=[0:w-1]$. Following established methods\citep{Singh2005, Krener2009, Cellini2023}, we empirically approximate the Jacobian by applying small perturbations ($\epsilon = 10^{-5}$) to each initial state variable $x_{i, 0}$, one at a time, in the positive ${x}^+_{i,0}$ and negative ${x}^-_{i,0}$ directions. We simulate the system using the inputs $\vect{u}_k$ calculated in step 3 (they are not recalculated for the perturbed trajectory), and then compute the difference in the corresponding change in sensory measurements over time $\Delta \vect{Y}_{\timewindow}^i = \vect{Y}_{\timewindow}^{i+} - \vect{Y}_{\timewindow}^{i-}$  (\autoref{fig:figure_1_d}). The end result for each state perturbation is a column vector of length $p \timewindow$, corresponding to $p$ measurements over $\timewindow$ time steps. We set a window size of $\timewindow=5$ steps for all simulations in this paper (see tips and tricks below for advice on selecting $\timewindow$). Repeating these steps for each state variable and stacking the resulting column vectors side-by-side yields the full empirically determined Jacobian, $\Delta \vect{Y}_{\timewindow} / \Delta \vect{x}_0=\Delta \vect{Y}_{\timewindow} / 2\epsilon$, which is equivalent to the empirical observability matrix $\mathcal{O}$ (\autoref{fig:figure_1_e}). 

This empirical approach can lead to misleading values if the measurement function has any discontinuities, such as angular wrapping issues. For our simulations, all angular measurement variables were first unwrapped to ensure that $\Delta \vect{Y}$ did not have any discontinuities. An alternative for angular measurements is to replace them with the sine and cosine of each angular measurement. 

\subsubsection*{5. Compute the Fisher information matrix and its inverse}
To compute observability levels for individual state variables in physically interpretable units, we construct the Fisher information matrix (\autoref{fig:figure_1_f}) as
\begin{equation} \label{equ:fisher}
    \mathcal{F} = \mathcal{O}^{\intercal} \mathcal{R}^{-1} \mathcal{O},
\end{equation}
where $\mathcal{R}$ is a block diagonal matrix of the measurement noise covariance matrices at each time step \citep{Boyacoglu2024}. In our simulations, we set $\mathcal{R} = 0.1 \cdot I_{(p \timewindow) \times (p \timewindow)}$, where $p$ is the number of measurements and $\timewindow$ is the sliding window size, assuming a noise variance of \SI{0.1}{rad} for angular measurements, \SI{0.1}{\meter\per\second} for airspeed and ground speed magnitude measurements, and \SI{0.1}{\per\second} for optic flow measurements. The absolute noise properties of these measurements do not have a large impact on our results because larger or smaller values of $\mathcal{R}$ simply scale $\mathcal{F}$; however, relative changes in $\mathcal{R}$ for different measurements would have an effect.
For the deterministic dynamics we consider, the process noise $Q$ is $0$ (see the following section for incorporating nonzero $Q$). The Fisher information matrix captures the amount of information that a known random variable, in our case the measurements $\vect{Y}_w$, carries about unknown parameters, in our case the initial state $\vect{x}_0$. The Cram\'{e}r--Rao bound states that the inverse of the Fisher information matrix $\mathcal{F}^{-1}$ sets the lower bound on the error covariance of the state estimate \citep{Cramer1946}. Thus, the diagonal elements of $\mathcal{F}^{-1}$ tell us the minimum possible variance in the error of any unbiased estimate for each state variable, inversely correlating with observability (\autoref{fig:figure_1_g}).
In cases where there is at least one unobservable state variable, $\mathcal{F}$ will not be invertible. Therefore, to compute $\mathcal{F}^{-1}$ we add a small regularization term to ensure invertibility \citep{chernoff1953locally}:
\begin{equation} \label{equ:fisher_inverse_limit_main}
    \mathcal{F}^{-1} =  \lim_{\lambda\to 0^+} {\tilde{\mathcal{F}}}^{-1} =  \lim_{\lambda\to 0^+} \big[ \mathcal{F} + \lambda I_{n \times n}\big]^{-1}.
\end{equation}
For unobservable state variables, this ``Chernoff inverse'' \citep{chernoff1953locally, rao1972generalized} yields infinity in the corresponding diagonal elements, whereas the more traditional Moore-Penrose pseudoinverse \citep{Penrose1955} yields misleading values that are often small when they should be very large to provide a correct interpretation with the Cram\'{e}r--Rao bound. Note that the regularization need not be limited to the form $\lambda I_{n\times n}$; a more general regularization of the form $\lambda J$, where $J$ is any $n\times n$ positive definite matrix, is also valid, though for simplicity we set $J = I_{n\times n}$. For computational efficiency, we use a small real value for $\lambda$ rather than taking the limit, which places an upper bound on the minimum error covariance equal to $1/\lambda$. Thus, $\lambda$ must be chosen carefully so as not to inject too much ``artificial" observability into the system while still avoiding numerical instabilities associated with computing a numerical inverse. We set $\lambda = \SI{1e-6}{}$ for all calculations.

\subsubsection*{6. Perform coordinate transformation}
The coordinates of both $\mathcal{O}$ and $\mathcal{F}$ are principally defined in the state space $\mathcal{X}$, which means that $\mathcal{F}^{-1}$ only provides information with respect to the individual state variables in $\vect{x}$. However, in many cases, it is beneficial to evaluate observability in a set of new coordinates $\vect{z}$. For instance, a system may be defined in Cartesian coordinates, but the polar representation of the state may be the most relevant for observability analysis. The system itself could be redefined in the new coordinates $\vect{f}(\vect{z}_k,\vect{u}_k)$ and the observability analysis pipeline could be repeated. However, this process is often tedious for complex systems, not computationally efficient, and is not practical for numerical solvers that rely on a specific coordinate frame. We incorporate a simple method to transform the coordinates of $\mathcal{O}$ (or $\mathcal{F}$) from an arbitrary state space coordinate basis $\vect{x}$ to a new basis $\vect{z}$ that bypasses these drawbacks and can be appended post hoc after computing $\mathcal{O}$ with respect to $\vect{x}$. 

First, the new basis $\vect{z}$ is defined as a transformation of $\vect{x}$ such that $\vect{z} = T(\vect{x})$, where $T(\vect{x})$ is a diffeomorphism that preserves the original information content in $\vect{x}$ \citep{VanTrees2001}. Because $\mathcal{O}$ fundamentally represents the Jacobian ${\partial \vect{Y}}/{\partial \vect{x}}$ in continuous time or ${\Delta \vect{Y}}/{\Delta \vect{x}}$ in discrete time, we can compute a coordinate transformation as the Jacobian of  $\vect{x}$ with respect to  $\vect{z}$, evaluated at the initial state  $\vect{z}_0=T\vect{x}_0$
\begin{equation} \label{equ:observability_transform}
\left. \frac{\partial \vect{x}}{\partial \vect{z}}\right|_{\vect{z} = \vect{z}_0}
= \left. \left( \frac{\partial T(\vect{x})}{\partial \vect{x}} \right)^{-1}\right|_{\vect{x} = \vect{x}_0}
\end{equation}
and transform the coordinates as
\begin{equation} \label{equ:observability_matrix_transform}
    \mathcal{O}_{\vect{z}} 
    =  \left.  \frac{\partial \vect{Y}}{\partial \vect{z}} \right|_{\vect{z} = \vect{z}_0}
    = \frac{\partial \vect{Y}}{\partial \vect{x}} \left. \frac{\partial \vect{x}}{\partial \vect{z}}\right|_{\vect{z} = \vect{z}_0}
    = \mathcal{O}_{\vect{x}} \left. \frac{\partial \vect{x}}{\partial \vect{z}}\right|_{\vect{z} = \vect{z}_0}.
\end{equation}
In practice,  $\partial \vect{x}/\partial \vect{z}$ can be computed numerically or symbolically and evaluated at the initial state $\vect{z}_0$. In some cases, it may be necessary to perform multiple coordinate transforms to evaluate the observability of new state variables that may not jointly satisfy the diffeomorphism/information requirement in a single state space.

\subsubsection*{7. Reveal active sensing motifs by correlating state dynamics with observability}
Finally, we repeat \textit{Steps} 1-6 in sliding time windows, that is, we use the same window size $\timewindow$ but with initial conditions that start progressively later in time along the nominal trajectory. For each window, we construct $\mathcal{O}$ and $\mathcal{F}$, optionally perform a coordinate transform, and calculate the minimum error variance for each state variable from ${\tilde{\mathcal{F}}}^{-1}$. This yields a time-series of minimum error variance along the state trajectory (\autoref{fig:figure_1_h}--\autoref{fig:figure_1_j}), which allows us to evaluate the temporal patterns of observability as a function of state variables and available sensory measurements.

In a toy example for the state trajectory in \autoref{fig:figure_1_a}, we can see that changes in heading angle lead to an order of magnitude decrease in the minimum error variance---equivalent to an increase in observability---for the first state variable $x_1$, compared to a straight trajectory (\autoref{fig:figure_1_i}, \autoref{fig:figure_1_h}). This result is based on measurements from three sensors $[y_1, y_2, y_3]$. However, different observability patterns emerge when considering other state variables and sensor combinations. 

For instance, the second state variable $x_2$ is observable during acceleration/deceleration and turns offset to the heading angle, but requires sensors $y_2$ to $y_p$, and does not require sensor $y_1$ (\autoref{fig:figure_1_j}, \autoref{fig:figure_1_h}). This illustrates that active sensing motifs can be tuned to the sensory measurements that are available and the state variable of interest. A critical point is that each state variable can have different observability properties, and thus distinct active sensing motifs. This approach allows us to investigate fundamental relationships between state variables, sensors, and active sensor motion. 

\subsection*{Additional considerations for observability calculations}

\textbf{Interpreting minimum error variance for circular variables}

The minimum error variance (MEV) provided by BOUNDS requires careful interpretation for circular variables. The value returned by BOUNDS is a variance measure that assumes that the underlying variable is linear. To correctly interpret this value for a circular variable, we can transform it into a circular variance that corresponds to a wrapped normal distribution with a standard deviation equal to $\sqrt{MEV}$. This circular variance can be found according to the following equation:
\begin{equation}
    \operatorname{Var_{circ}}=1-\exp{(-MEV/2)},
\end{equation}
where $MEV$ is the minimum error variance given BOUNDS, e.g. $\mathcal{F}_i^{-1}$.  

\subsubsection*{Observability for non-contiguous time windows}
In some cases it is important to consider the times at which measurements are collected when evaluating active sensing motifs. For some sets of measurements, information must be collected continuously throughout a motif to increase observability (\autoref{fig:figure_supp_timestep_a}). However, other sets of measurements may allow information to be considered only before and after the motif, with a similar increase in observability to continuously collecting information (\autoref{fig:figure_supp_timestep_b}). This is an especially important consideration when dealing with systems where sensor noise may be correlated with a motif. For instance, translational optic flow cues may be noisier or less reliable when turning due to the influence of rotational optic flow.

\subsubsection*{Constructability and inclusion of model uncertainty}
Our observability framework quantifies the minimum error variance for the value of an initial state variable given \textit{subsequent} measurements, i.e. an error bound on $\hat{x}_{i,k}$ given $\mathbf{Y}_{k:k+\timewindow}$. In state estimation, we are generally interested in building state estimates given \textit{preceding} measurements, that is, we want to find $\hat{x}_{i,k}$ given $\mathbf{Y}_{k-\timewindow:k}$. To quantify the error bound for this final state, it would be more accurate to use one of observability's dual principles, namely constructability \citep{Boyacoglu2024}, also referred to as the posterior Cramér-Rao bound \citep{tichavsky1998posterior}, or observability of the final state \citep{sontag1990}. We chose to analyze observability, rather than constructability, because it allowed us to use an initial state perturbation approach for empirically calculating the observability matrix $\mathcal{O}$. This perturbation approach is efficient, and works seamlessly with non-closed-form models.  For the purpose of discovering active sensing motifs, both approaches will provide equivalent insight. In cases where a quantification of the error bounds for the final state is necessary, our calculations of $\mathcal{F}$ can be replaced with an iterative equation that uses a discretized time-varying approximation of the continuous-time dynamics \citep{Boyacoglu2024}. Taking this approach would also allow for the inclusion of non-zero process noise covariance matrix $Q$ to understand how uncertainty of an agent's internal model affects the observability (or constructability) of state variables of interest. We note that for the relatively short time windows and small model covariances ($Q\approx0$) that we considered, the observability and constructability will generally be strongly correlated, allowing us to use our empirical observability calculations as a proxy for constructability. For our demonstration of the AI-KF we circumvented this issue by empirically establishing the error bound of a data-driven estimator by analyzing its error variance directly. 

\subsubsection*{Tips and tricks}
In general, our method is appropriate for any system, as long as the system can be simulated. However, it is critical to ensure that none of the state variables have discontinuities in the state space of the simulation. Otherwise, artifacts can manifest as extremely large values in the observability matrix, which corrupt the computation of the Fisher information matrix. For instance, when dealing with polar variables, such as angles, it is crucial to ensure that the state perturbations required to construct the observability matrix do not lead to angle wrapping discontinuities. Magnitude variables must not be perturbed to be negative if using central/backwards difference methods to calculate a Jacobian, as magnitudes are defined to always be positive. This would have the same effect in the observability matrix as applying a $180^{\circ}$ perturbation to the associated angular quantity. It is possible, and recommended, to represent a single angular measurement as the sine and cosine components of the unit vector associated with the angle---this can mitigate the chance for wrapping discontinuities. It is also important to ensure that all measurement functions are always defined. If a measurement consists of a fraction such as $y_1/y_2$, then $y_2=0$ would lead to an undefined measurement which breaks observability assumptions. 

When considering the observability associated with many different combinations of sensors, it is best practice to compute the observability matrix once for all the potential state variables and sensors, and for the largest time window to be considered. Then one can pull out smaller observability matrices corresponding to whatever combinations of state variables, sensors, and time windows for further analysis (Fisher information, etc.). This is in contrast to performing many simulations to reconstruct a new observability matrix for each unique combination, which effectively contains the same information. This approach avoids cumbersome computations and makes iterative investigation of active sensing motifs much faster, although this is dependent on the speed of the simulation required to construct the observability matrix.

\subsubsection*{Hyper-parameters}
BOUNDS has two primary hyper-parameters, 1) the regularization parameter $\lambda$ (Eq. \ref{equ:fisher_inverse_limit_main}), and 2) the window size $\timewindow$ over which to analyze the observability. It is best practice to choose $\lambda$ to be large enough to numerically invert the Fisher information matrix, but small enough to not distort the total information. In general, a good guideline is to choose $\lambda$ to be less than the smallest nonzero eigenvalue of the Fisher information matrix \citep{Horn2012}. Alternatively, one could symbolically compute the limit in Eq. \ref{equ:fisher_inverse_limit_main}, however, this can be computationally expensive and can yield infinite values corresponding to unobservable state variables. For $\timewindow$, we recommend choosing a window size roughly equal to the slowest motif of interest. 

\subsection*{Flying agent dynamical models}
\subsubsection*{Kinematic model}
The primary model employed for observability analysis (see \autoref{fig:figure_2}) was a 3D kinematic model for a flying agent. We use the word ``kinematic" here to describe a model that does not rely on force and torque inputs, but rather inputs that can be measured kinematically, like accelerations and angular rates. We chose to use a kinematic model because it is more agnostic to the diverse range of dynamics of flying agents, and may provide more general conclusions. Although we use a continuous-time model to describe the underlying dynamics, the dynamics were effectively converted to discrete-time when solving the differential equations. The model is defined in state space as
\begin{equation} \label{equ:model_kinematic_3D}
\vect{\dot{x}} =
    \begin{bmatrix} 
        \dot{x} \\
        \dot{y} \\
        \dot{z} \\
        \dot{v}_x \\
        \dot{v}_y \\
        \dot{v}_z \\
        \dot{\phi} \\
        \dot{\theta} \\
        \dot{\psi} \\
        \dot{w} \\
        \dot{\zeta}\\
    \end{bmatrix}
    =
    \begin{bmatrix}
        v_x \cos(\psi) - v_y \sin(\psi) \\
        v_x \sin(\psi) + v_y \cos(\psi) \\
        v_z \\
        u_z k_z\cos(\phi) \sin(\theta) - C a_x + v_y \dot{\psi} \\
        -u_z k_z\sin(\phi) - C a_y - v_x \dot{\psi} \\
        -u_z k_z \cos(\phi) \cos(\theta) - C v_z + \mathscr{g} \\
        u_{\phi} k_{\phi}\\
        u_{\theta} k_{\theta}\\
        u_{\psi} k_{\psi} \\
        u_{w} \\
        u_{\zeta}\\
    \end{bmatrix}.
\end{equation}
The model describes the velocity of the agent in the body-level North-East-Down (NED) coordinate system. $[v_x, v_y]$ define translational velocities in the $XY$ plane that rotate with the agents yaw heading $\psi$ and $v_z$ describes the global $Z$ velocity. Note that the derivatives of $[v_x, v_y]$ include the derivative of the rotating reference frame. The body-level frame was chosen primarily because it simplifies the control of trajectories. The position of the agent $[x, y, z]$ was defined in the global frame to more easily visualize trajectories. In general, the choice of reference frame does not affect observability results and only serves to simplify some of the intermediate steps in our framework.

The model has four inputs for flight control. The acceleration in the body frame $u_z$  serves to counteract acceleration from gravity $\mathscr{g}$ (note: we use the symbol $g$ to refer to ground speed, and $\mathscr{g}$ for acceleration from gravity, and do not refer to gravity anywhere except this line and the $\dot{v}_z$ dynamics). By changing roll $\phi$ and pitch $\theta$, the agent can redirect $u_z$ and control its velocity vector. The inputs $[u_{\phi}, u_{\theta}, u_{\psi}]$ are angular rates that allow the agent to control roll, pitch, and yaw. Most flying systems use a control allocation\citep{Johansen2013} approach to transform these basic inputs into control commands for individual actuators, like motor or tilter commands.  However, this level of detail is unnecessary for the purpose of observability analysis and only serves to complicate the control of the agent. Thus, we do not model the actuation associated with these inputs. We augmented the state vector with motor calibration coefficient states for each input $[k_z, k_{\phi},  k_{\theta},  k_{\psi}]$, which were all set equal to one. These augmented states allowed us to consider cases where the agent did not know part of its input dynamics.

We model drag as proportional to the 2D apparent airflow $[a_x, a_y]$ through a drag coefficient $C$ to allow the ambient wind to influence the acceleration of the agent---note that this is a kinematic drag and not a force. The apparent airflow is defined as
\begin{equation} \label{equ:model_apparent_windl}
    \begin{bmatrix} 
        {a_x} \\
        {a_y} \\
    \end{bmatrix}
    =
    \begin{bmatrix}
        -v_{x} + w \cos{(\psi-\zeta)} \\
        -v_{y} - w \sin{(\psi-\zeta)} \\
    \end{bmatrix},
\end{equation}
where $w$ is the ambient wind magnitude and $\zeta$ is the ambient wind direction in the $XY$ plane (\autoref{fig:figure_2_b},--\autoref{fig:figure_2_c}). To model time-varying wind, we added inputs to control the ambient wind magnitude $u_w$ and direction $u_{\zeta}$.

\subsubsection*{Dynamical model}
To test the sensitivity of our observability analysis to modeling choices, we also employed a model with more realistic 3D dynamics \cite{Beard2008} (\autoref{fig:figure_supp_dynamic}). This model is distinct from the kinematic model described above in that it uses forces and torques as inputs and considers physical parameters like mass $m$, inertia $I_{\bullet}$, and drag forces $C_{\bullet}$ about each axis. The states remained the same as in \autoref{equ:model_kinematic_3D}, but we added state variables for angular velocity $[\omega_x, \omega_y, \omega_z]$.
\begin{equation} \label{equ:drone_model}
{\renewcommand{\arraystretch}{1.5}
\vect{\dot{x}} =
    \begin{bmatrix} 
        \dot{x} \\
        \dot{y} \\
        \dot{z} \\
        \dot{v}_x \\
        \dot{v}_y \\
        \dot{v}_z \\
        \dot{\phi} \\
        \dot{\theta} \\
        \dot{\psi} \\
        \dot{\omega}_x \\
        \dot{\omega}_y \\
        \dot{\omega}_z \\
        \dot{w} \\
        \dot{\zeta}\\
    \end{bmatrix}
    =
    \begin{bmatrix}
        v_x \cos(\psi) - v_y \sin(\psi) \\
        v_x \sin(\psi) + v_y \cos(\psi) \\
        v_z \\
        \frac{1}{m} (u_z \cos(\phi) \sin(\theta) - C a_x) + v_y \dot{\psi} \\
        \frac{1}{m} (-u_z \sin(\phi) - C a_y) - v_x \dot{\psi} \\
        \frac{1}{m} (-u_z \cos(\phi) \cos(\theta) - C v_z + m\mathscr{g}) \\
        \omega_x + \tan(\theta) (\omega_y \sin(\phi) + \omega_z \cos(\phi)) \\
        \omega_y \cos(\phi) - \omega_z \sin(\phi) \\
        \omega_y \frac{\sin(\phi)}{\cos(\theta)} - \omega_z \frac{\cos(\phi)}{\cos(\theta)}  \\
        \frac{u_{\phi}}{I_x} + \frac{I_y - I_z}{I_x} \omega_y \omega_z \\
        \frac{u_{\theta}}{I_y} + \frac{I_z - I_x}{I_y} \omega_x \omega_z \\
        \frac{u_{\psi}}{I_z} + \frac{I_x - I_x}{I_z} \omega_x \omega_y \\
        \dot{w} \\
        \dot{\zeta}\\
    \end{bmatrix}}.
\end{equation}

Parameter values were chosen based on those measured or estimated for our physical quadcopter system (\autoref{table:model_params}). For observability calculations, we included the mass, inertia, and drag parameters in \autoref{table:model_params} as auxiliary state variables. This allowed us to specifically investigate observability when the agent does not have explicit knowledge of its own parameters. 
\begin{table}[h!] 
\centering
\fontsize{9.0pt}{9.0pt}\selectfont
\renewcommand{\arraystretch}{1.5} 
\begin{tabular}{llc}
\hline
\textbf{Parameter} & \textbf{Value} \\ \hline
mass ($m$) & \SI{2.529}{\kg} \\ \hline
roll inertia ($I_x$) & \SI{0.040}{\kg\meter\squared} \\ \hline
pitch inertia ($I_y$) & \SI{0.040}{\kg\meter\squared} \\ \hline
yaw inertia ($I_z$) & \SI{0.046}{\kg\meter\squared} \\ \hline
translational damping ($C$) & \SI{0.1}{\kg\per\second} \\ \hline
\end{tabular}
\caption{\textbf{Quadcopter model parameters}. All values computed for the physical quadcopter system in \autoref{fig:figure_5_a}.}
\label{table:model_params}
\end{table}

\subsubsection*{Measurement model}
The measurements of our models were chosen depending on which set of measurements we were investigating (\autoref{fig:figure_2}). However, all measurements were defined as a function of the state variables, and chosen from the set
\begin{equation} \label{equ:model_dynamic_3D}
   { \renewcommand{\arraystretch}{1.5}
    \vect{h}(\vect{x}) = 
    \begin{bmatrix} 
        \psi \\
        \beta \\
        \gamma \\
        \eta \\
        g \\
        a \\
        r \\
        q
    \end{bmatrix}
    =
    \begin{bmatrix} 
        \psi \\
        \atantwo (v_{y}, v_{x}) \\
        \atantwo (a_{y}, a_{x})-2\pi \\
        \atantwo (\dot{v}_{y}, \dot{v}_{x}) \\
        \sqrt{{v_{x}^2 + v_{y}^2}} \\
        \sqrt{{a_{x}^2 + a_{y}^2}} \\
        g / z \\
        \sqrt{{(\dot{v}_{x}-v_y\dot{\psi})^2 + (\dot{v}_{y}+v_x\dot{\psi})^2}} 
    \end{bmatrix}}.
\end{equation}
\noindent The $-v_y\dot{\psi}$ and $+v_x\dot{\psi}$ terms are present to cancel out the terms from \autoref{equ:drone_model} that describe the derivative of the rotating reference frame, as a body mounted accelerometer would not pick up these terms. The $-2\pi$ term for $\gamma$ is necessary because $\gamma$ is defined in \ref{fig:figure_2} with the vector pointed towards the agent, rather than away.

\subsubsection*{Simulation and control}
We simulated our continuous-time dynamics models in Python with an ordinary differential equation (ODE) solver \cite{Fiedler2023}. As opposed to applying the ODE solver once across the full time-series, we applied it to each discrete time step independently while holding the inputs constant at each step. This was done as opposed to interpolating inputs between time steps, as interpolation between steps can cause issues reproducing simulations starting from an arbitrary point along a trajectory due to the way numerical solvers operate. Our discrete time approach ensured that we could start from any given point along a trajectory and maintain the exact simulation thereafter, which was necessary to construct observability matrices in sliding windows.

We used model predictive control (MPC) with full state feedback to precisely control flight trajectories through our model (\autoref{equ:drone_model}) (\autoref{fig:figure_1_b}). In principle, other control methods could work, but MPC is a good general choice for controlling nonlinear systems. We first prescribed---or took from measured trajectories---a desired set-point for $v_{x}$, $v_{y}$, and $\psi$ over time (\autoref{fig:figure_1_a}), then used an open-source MPC toolbox \citep{Fiedler2023} to find the optimal control inputs to track the set-point variables. Our MPC cost function penalized the squared error between each of our set-point series while adding an additional penalty for each control input to ensure smoother force and torque inputs. Specifics can be found in our publicly available code.

\subsection*{Artificial neural networks}
\subsubsection*{Wind direction estimator}
We designed a feed-forward artificial neural network (ANN) for regression to estimate wind direction \autoref{fig:figure_3_a}. There were 12 input neurons consisting of the heading $\psi$, apparent airflow angle $\gamma$, and optic flow angle $\beta$ in the time window $\timewindow=4$. We used three hidden layers with 64 neurons and an output layer with a single neuron to estimate wind direction $\zeta$. All neurons were fully connected and used a rectified linear unit activation function, except for the output layer, which used a linear activation function. We also added a Gaussian noise layer after the input layer  ($\mu = 0$, $\sigma = 0.01$ rad), which was only active during training, to help the ANN deal with noise in the measurements. We built and trained our network using TensorFlow \citep{Abadi2016} and Keras \citep{Chollet2015}. We designed a custom loss function to deal with the circular nature of the output variable:
\begin{equation*} \label{equ:loss_function}
        f(\zeta, \check{\zeta}) = {(\sin(\zeta) - \sin(\check{\zeta}))}^{2} + {(\cos(\zeta) - \cos(\check{\zeta}))}^{2},
\end{equation*}
where $\zeta$ is the true wind direction of the training data and $\check{\zeta}$ is the ANN output (predicted $\zeta$). This function penalizes the squared error between the sine and cosine components of $\zeta$ and $\check{\zeta}$.

We generated a dataset of 40,000 simulated trajectories to train and test the wind direction estimator ANN (\autoref{fig:figure_3_b}). We ensured that our dataset consisted of a diversity of trajectory types by randomly choosing parameters from a uniform distribution between a set range of values for each of the following properties
\begin{flalign} \label{equ:ann_data}
    & w = [0, 2] \nonumber \\
    & \zeta = [-\pi, \pi] \nonumber \\
    & \psi = [-\pi, \pi] \nonumber \\
    & \dot{\psi} = [-0.11, 0.11] \nonumber \\
    & \ddot{\psi} = [-2.08, 2.08] \nonumber \\
    & v_{x} = [0.05, 5] \nonumber \\
    & v_{x, \timewindow} / v_{x, 0} = [0.05, 2.0] \nonumber \\
    & v_{y} = [-0.2, 0.2] \nonumber \\
    & \dot{v}_{y} = [-0.1, 0.1], \nonumber
\end{flalign}
where $v_{x, \timewindow} / v_{x, 0}$ represents an acceleration as the ratio of initial and final $v_{x}$. This was used instead of $\dot{v}_{x}$ to ensure flight was always generally directed forward and there was not backwards flight for small starting $v_{x}$ values. We applied a 80-20 train-test split to the data. Each trajectory was \SI{0.2}{\second} (21 discrete points). We augmented the measurements from each trajectory with $\timewindow=4$ prior time steps, so only the 17 points with $\timewindow$ prior points available were used in training/testing. The training and testing data split into equally sized bins, once sorted by observability and once sorted randomly (\autoref{fig:figure_3_c}). We trained and tested multiple ANNs on these binned datasets (\autoref{fig:figure_supp_ANN_percentile}). We trained each ANN for 10000 epochs with a batch size of 4096. Specific details can be found in our publicly available code.

\subsubsection*{Altitude estimator}
The altitude estimator ANN was designed similarly to the wind direction estimator ANN (\autoref{fig:figure_supp_altitude_ANN_a}). The ANN had 40 input neurons consisting of the forward optic flow $r_x$ and forward acceleration $\dot{v}_x$ in the time window $\timewindow=20$. Like the wind direction estimator, we used three hidden layers with 64 neurons and an output layer with a single neuron to estimate altitude $z$.

We generated a dataset of 2,000 simulated trajectories to train and test the altitude estimator ANN. Each trajectory had a constant altitude between 0 and 20 meters chosen randomly from a uniform distribution. To generate a wide range of forward accelerations, we varied the forward velocity via a sum-of-sines signal with frequency components between 0.1 and 0.9 Hz,  amplitudes between $-15$ and $15  \text{ ms}^{-1}$, phase between $-\pi/2$ and  $\pi/2$ , and an offset between $-10$ and $10  \text{ ms}^{-1}$. We applied a 80-20 train-test split to the data. Each trajectory was \SI{11}{\second} (111 discrete points). We augmented the measurements from each trajectory with $\timewindow=20$ prior time steps, so only the 91 points with $\timewindow$ prior points available were used in training/testing.

\subsection*{Augmented Information Kalman Filter}
\subsubsection*{Implementation details}
We used the MATLAB implementation of an Unscented Kalman Filter (which uses the utilize square-root factorization of covariance matrices for improved numerical stability)  \citep{Wan2006} for all Augmented Information Kalman Filter simulations. The Unscented transform parameters were set to $\alpha = 10^{-3}, \beta = 1, \kappa=0$. The initial state and covariance estimates, process noise covariance, and measurement noise were varied (\autoref{fig:figure_4}). The baseline noise variance for the optic flow measurement was set to $10^{-3}$. Specifics can be found in our publicly available code.

\subsubsection*{Eliminating double counting of information}
After the AI-KF has converged, and during active sensing maneuvers yielding trustworthy (i.e. highly observable; low $\operatorname{Var}(\tilde{\check{x}}_i)$) values for the data driven estimates, $\check{x}_{i}$ and $\hat{x}_{i-1}$ will not necessarily be independent. This breaks assumptions of the Kalman framework and would lead to an artificially low error variance for the estimates. This can be resolved through a small modification of $\mathcal{R}^{\prime}$ with the following steps. 

\begin{enumerate}
    \item Determine two estimates: a naive KF estimate $\hat{\ubar{x}}_i$ that ignores the augmented terms (i.e. by setting $\operatorname{Var}(\tilde{\check{x}}_i)$ to a large number approaching infinity, and the off-diagonal elements to zero), and the data-driven estimate $\check{x}_i$. 
    \item Calculate the difference between these two estimates: 
\begin{equation}
    d_i=\hat{\ubar{x}}_{i}-\check{x}_{i}.
\end{equation}
    \item Define a ``relevance ratio": 
\begin{equation}
r_i=\frac{d_i^2}{\operatorname{Var}(\tilde{\check{x}}_i)}.
\end{equation}
When $r_i$ is small it means that the data-driven estimate carries little unique information, and should therefore be ignored. When $r_i$ is large, the data-driven estimate carries new information and should be used to augment the KF. 
    \item Define a throttling function that can be used to suppress the augmented measurement,
\begin{equation}
    g(r_i) = 1 + \frac{c}{r_i + \epsilon}, \quad c > 0, \quad \epsilon > 0,
\end{equation}
where $c$ is a tunable hyperparameter that controls the strength of suppression, and $\epsilon$ prevents division by zero and sets a ceiling on $g(r_i)$. When the data-driven estimate carries no new information $g(r_i)$ is very large (approaching $c/\epsilon$). When the data-driven estimate does carry new information $g(r_i)$ approaches $1$. 
    \item Multiply the $\operatorname{Var}(\tilde{\check{x}}_i)$ in $\mathcal{R}^\prime$ by $g(r_i)$, and divide the off-diagonal elements by $\mathcal{R}^\prime$ by $g(r_i)$. 
\end{enumerate}

This framework ensures that when the data-driven estimate $\check{x}_{i}$ is both accurate (as indicated by $\operatorname{Var}(\tilde{\check{x}}_i)$) and different from the naive KF estimate (i.e. $|d_i|\gg0$, the AI-KF takes full advantage of the augmented information. However, when the KF has converged and produces estimates that are statistically indistinguishable to the data-driven estimates, the AI-KF automatically relaxes to a naive KF, eliminating the possibility of double counting information.

\subsection*{Outdoor experiments}
We collected data from a quadcopter (DJI Matrice 300 RTK) flying outdoors over relatively flat terrain near Reno, NV (\autoref{fig:figure_5_a}). Acceleration data was measured by the onboard accelerometer and ventral camera images were collected at 60 Hz by a ventrally mounted Raspberry Pi camera. Optic flow was computed offline using an implementation of Dense Inverse Search in OpenCV (Python) and low-pass filtered with a cutoff frequency of 1 Hz \citep{opencv_library}. We scaled the optic flow with a constant scalar (approximately 0.057) to account for camera and lens properties such that optic flow was equivalent to the ratio of velocity over altitude. Ground truth data was measured by an onboard GPS. All flights were operated in velocity control mode, with manually specified accelerations/decelerations. 

\section*{Data and code availability}
\label{sec:Code}
We provide a Python package to implement BOUNDS (\href{https://github.com/vanbreugel-lab/pybounds}{https://github.com/vanbreugel-lab/pybounds}) which will be updated periodically. Specific data associated with this manuscript will be made public upon publication.

\section*{Funding}
This work was supported by a National Science Postdoctoral Fellowship in Biology to Benjamin Cellini (NSF 22-623). The work was also supported by the Air Force Office of Scientific Research (FA9550-21-0122) to FvB, the NSF AI institute in Dynamics (2112085) to FvB, NSF-EFRI-BRAID (2318081) to FvB, Sloan Research Fellowship (FG-2020-13422) to FvB, DEPSCoR (FA9950-23-1-0483) to FvB, and NIH (1R01NS136988) to FvB. This research was also supported in part by the NSF grant PHY-2309135 to the Kavli Institute for Theoretical Physics (KITP), which supported valuable discussions that led to substantial improvements of the paper. For the purpose of open access, the author has applied a CC BY public copyright license to any Author Accepted Manuscript version arising from this submission.

\section*{Acknowledgments}
We thank Christina May for productive conversations about applications of observability analysis on real data and feedback on \textit{pybounds} code, Stanley David Stupski for providing data for testing, J. Humberto Ramos for help developing the quadcopter dynamical model, and Natalie Brace and Aditya Nair for helpful comments on the manuscript.

\section*{Conflicts of interest}
The authors declare no conflict of interest.

\section*{Author contributions}
Conceptualization: B.C., B.B., F.v.B.;
Methodology: B.C., B.B., A.L., F.v.B.;
Software: B.C.;
Formal analysis: B.C.;
Investigation: B.C., F.v.B.;
Resources: F.v.B.;
Data curation: B.C., A.L.;
Writing - original draft: B.C., F.v.B.;
Writing - review \& editing: B.C., B.B., A.L., F.v.B.;
Visualization: B.C., F.v.B.;
Supervision: F.v.B.;
Project administration: F.v.B.;
Funding acquisition: B.C., F.v.B.

\newpage
\clearpage
\bibliography{main.bib}

\clearpage

\onecolumn{

\beginsupplement
\renewcommand\figurename{Supplementary Figure}
\renewcommand\tablename{Supplementary Table}

\section*{Supplementary Material}

\begin{figure*}[!th]
    \centerline{\includegraphics{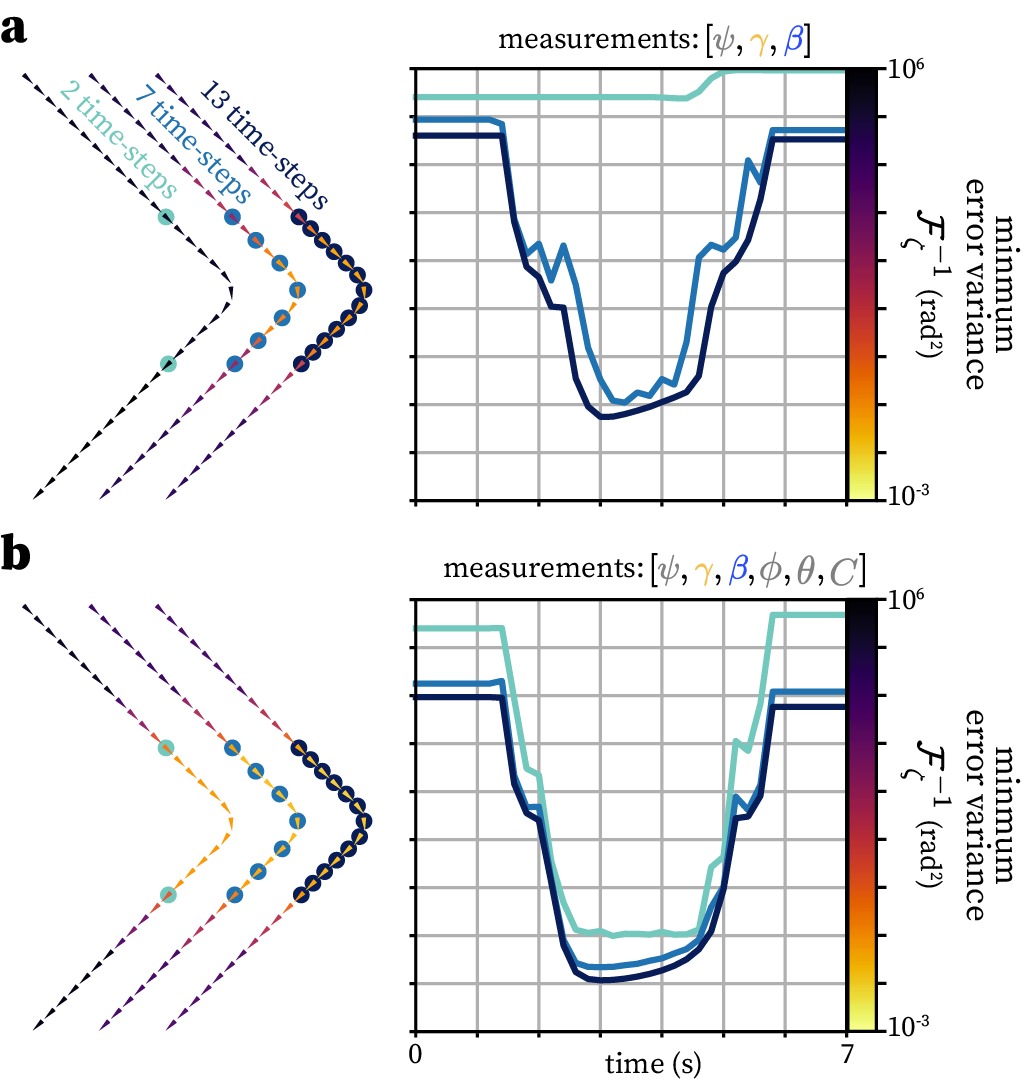}}
    \caption{\textbf{Observability for measurements at different times during an active sensing motif.}
    \textbf{a.} The observability of wind direction (calculated as in \autoref{fig:figure_2}) for the same trajectory, but considering measurements of heading ($\psi$), course direction ($\beta$) and apparent airflow angle ($\gamma$) only at specific times during the heading turn. Note that the turn does not increase observability when only considering measurement time steps before and after the turn (green) compared to using time steps throughout the turn (blues).
    \textbf{b.} Same as \textit{a} but with extra measurements of the agent's roll angle ($\phi$), pitch angle ($\theta$), and damping translational coefficient ($C$). Note that this set of measurements enables only two measurement time steps, before and after the turn, to increase observability.
    }
    \createsubpanels{figure_supp_timestep}{2}
    \label{fig:figure_supp_timestep}
\end{figure*}

\clearpage

\begin{figure*}[!th]
    \centerline{\includegraphics{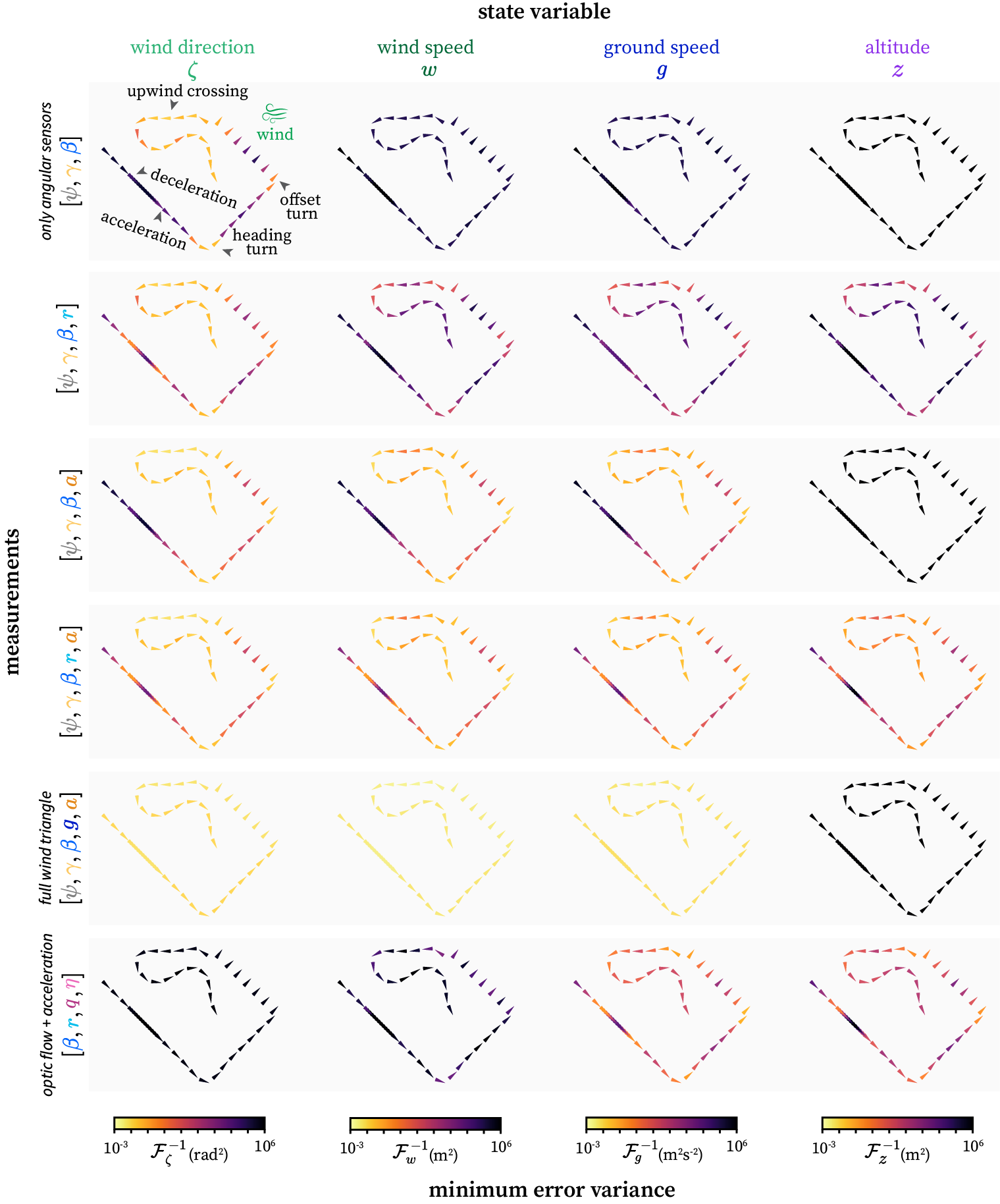}}
    \caption{\textbf{Active sensing motifs with a full dynamical drone model.}
    Same as \autoref{fig:figure_2} but using a dynamical model.
    }
    \label{fig:figure_supp_dynamic}
\end{figure*}

\clearpage

\begin{figure*}[!th]
    \centerline{\includegraphics{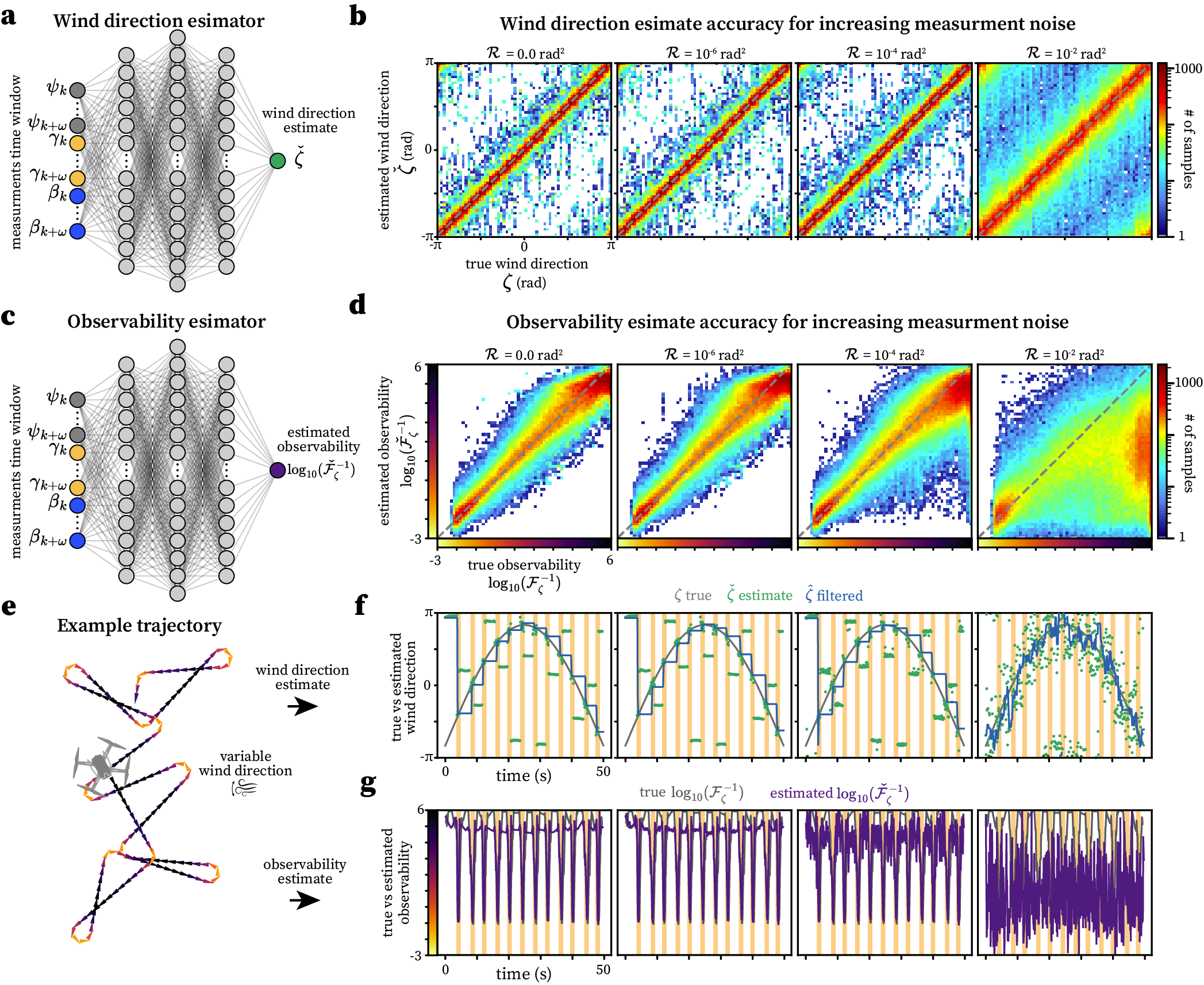}}
    \caption{\textbf{ANN Wind direction and wind direction observability estimator performance for varying measurement noise levels.}
    \textbf{a.} Wind direction artificial neural network (ANN) estimator (same as \autoref{fig:figure_3_c}).
    \textbf{b.} Wind direction ANN estimator testing performance as a function of the input (measurement) noise levels, where $\mathcal{R} = 10^{-4}$ indicates a noise covariance matrix of $\mathcal{R} = 10^{-4} \times I_{p \times p}$ for the ANN inputs. The gray dashed line indicates perfect performance.
    \textbf{c.} Wind direction observability estimator ANN estimator.
    \textbf{d.} Same as \textit{b.} but for the wind direction observability estimator performance.
    \textbf{e.} An example simulated trajectory (same as \autoref{fig:figure_3_f}) with variable wind direction.
    \textbf{f.} The raw wind direction ANN estimate for the trajectory in \textit{e.} (green) vs true wind direction (gray) for varying measurement noise levels. The ``observability-filtered" estimate (see \autoref{fig:figure_supp_observability_filter}) is shown in blue. The orange patches indicate the times of the observable turns during the trajectory.
    \textbf{g.} The wind direction observability ANN estimate for the trajectory in \textit{e.} (purple) vs true observability (gray) for varying measurement noise levels.
    }
    \label{fig:figure_supp_ANN_noise}
    \createsubpanels{figure_supp_ANN_noise}{7}
\end{figure*}

\clearpage

\begin{figure*}[!th]
    \centerline{\includegraphics{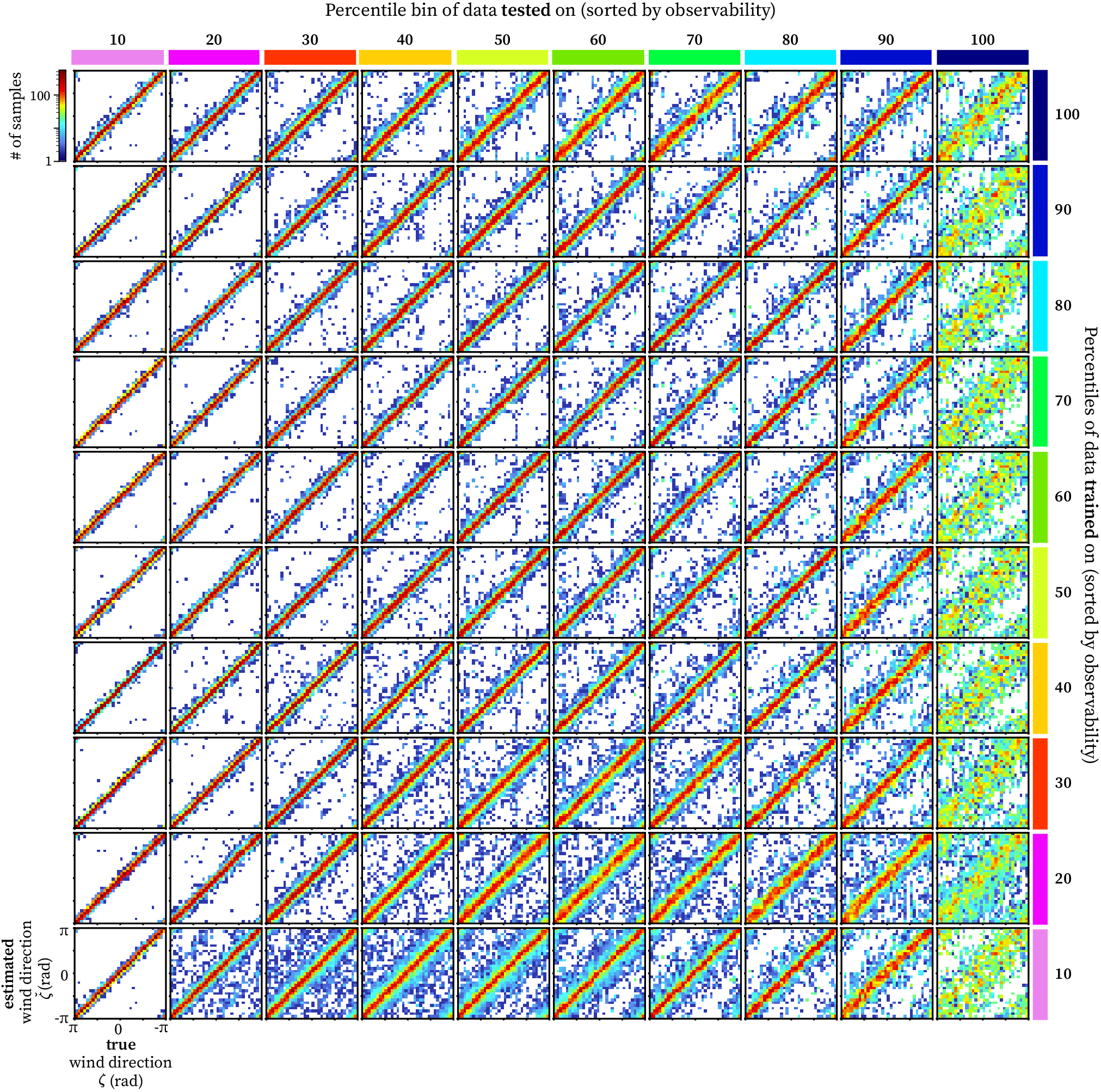}}
    \caption{\textbf{ANN Wind direction estimator performance.}
    Wind direction artificial neural network (ANN) estimator (see \autoref{fig:figure_3_c}) testing performance as a function of training and testing data with different observability levels. The dataset in \autoref{fig:figure_3_b} was split into 10 equally sized bins and an ANN was trained on 80\% of the data from each bin. The withheld 20\% of each bin was used to evaluate the performance. \autoref{fig:figure_3_c} shows the testing estimation error variance $\text{Var}(\zeta- \check{\zeta})$ for each bin corresponding to each panel here compared to ANNs trained on randomly split data. The ANN input (measurement) noise variance were all set to $10^{-4}$ rad, corresponding to a noise covariance of $\mathcal{R} = 10^{-4} \times I_{p \times p}$.
    }
    \label{fig:figure_supp_ANN_percentile}
\end{figure*}

\clearpage

\begin{figure*}[!th]
    \centerline{\includegraphics{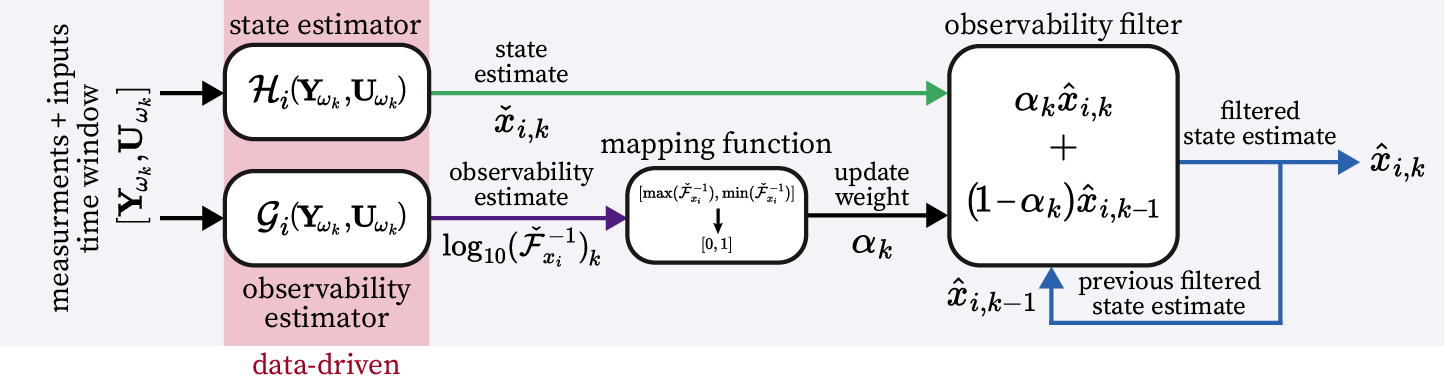}}
    \caption{\textbf{Observability-filter.}
    Overview of observability-informed state estimation framework. Artificial neural networks (ANN) estimate a state variable ${x}_{i,k}$ ($\mathcal{H}_i$) and its observability $\mathcal{F}^{-1}_{x_{i,k}}$ ($\mathcal{G}_i$) from available measurements $\vect{Y}_{\timewindow_k}$ and $\vect{U}_{\timewindow_k}$ inputs in the time window $\timewindow$. The state estimate $\check{x}_{i,k}$ is adaptively filtered based on the estimated observability $\check{\mathcal{F}}^{-1}_{x_{i,k}}$ to change more quickly when observability is high.
    }
    \label{fig:figure_supp_observability_filter}
\end{figure*}

\clearpage

\begin{figure*}[!th]
    \centerline{\includegraphics{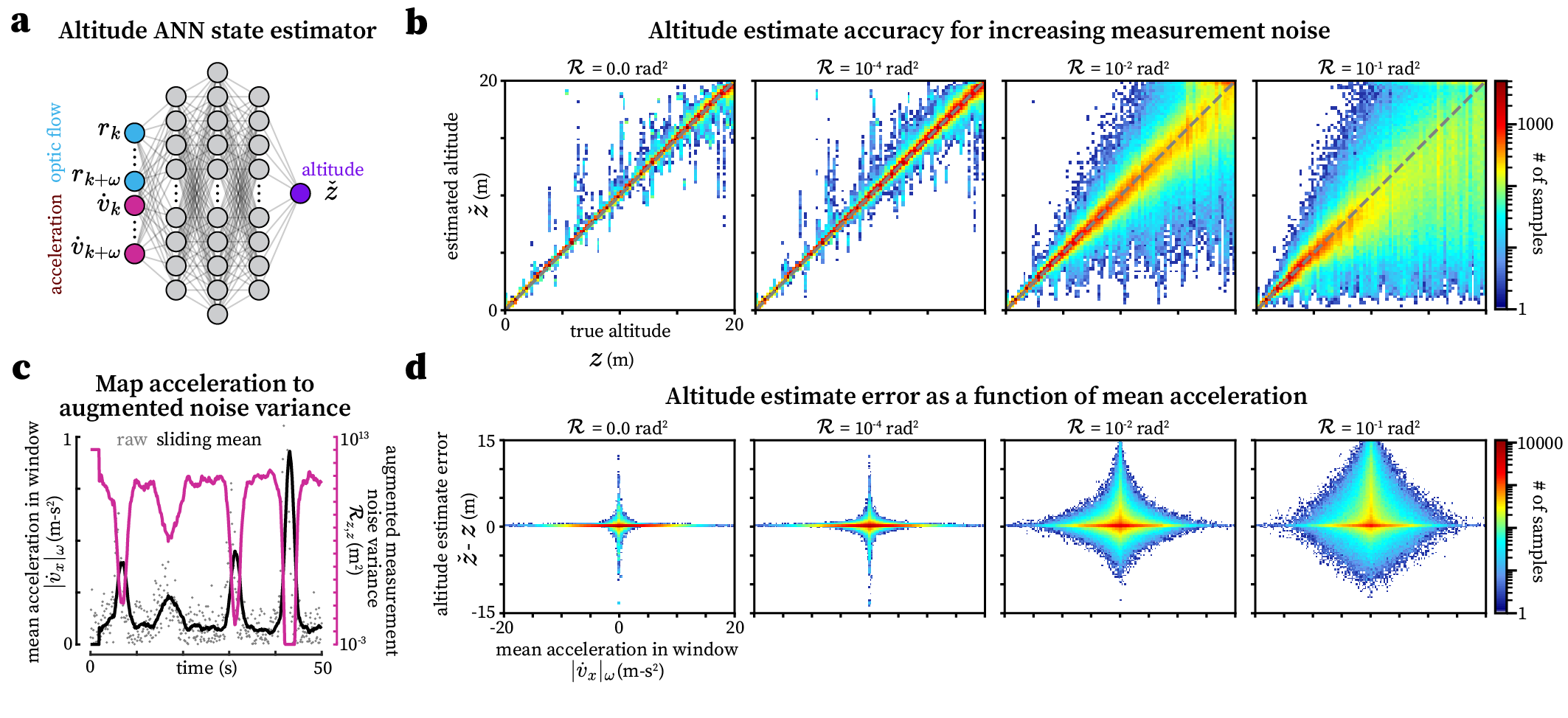}}
    \caption{\textbf{ANN altitude estimator.}
    \textbf{a.} Artificial neural network (ANN) that estimates altitude from measurements of forward optic flow ($r_x$) and forward acceleration ($\dot{v}_x$) over the time window $\timewindow$.
    \textbf{b.} Altitude ANN estimator testing performance as a function of the input (measurement) noise levels, where $\mathcal{R} = 10^{-4}$ indicates a noise covariance matrix of $\mathcal{R} = 10^{-4} I_{p \times p}$ for the ANN inputs. The gray dashed line indicates perfect performance.
    \textbf{c.} Illustration of mapping the mean forward acceleration ($\dot{v}_x$) in the window $\timewindow$ to the augmented noise variance $\check{\mathcal{R}}_{z,z}$ for the altitude estimate.
    \textbf{d.} Altitude ANN estimator testing performance as a function of the mean forward acceleration ($\dot{v}_x$) in the window $\timewindow$, as a function of the input (measurement) noise levels. Note that the shape of the heatmap illustrates the variance of the altitude estimates.
    }
    \createsubpanels{figure_supp_altitude_ANN}{4}
    \label{fig:figure_supp_altitude_ANN}
\end{figure*}

}

\end{document}